\documentclass[journal]{IEEEtran}
\pdfoutput=1
\usepackage{amsmath,amsfonts}
\usepackage{algorithmic}
\usepackage{algorithm}
\usepackage{array}
\usepackage[caption=false,font=normalsize,labelfont=sf,textfont=sf]{subfig}
\usepackage{textcomp}
\usepackage{stfloats}
\usepackage{url}
\usepackage{verbatim}
\usepackage{graphicx}
\usepackage{cite}
\usepackage{subeqnarray}
\usepackage{cases}
\usepackage{amssymb}

\usepackage{textcomp}
\usepackage{xcolor}
\usepackage{float}
\usepackage{multirow}
\usepackage{booktabs}
\usepackage{threeparttable}
\newtheorem{proposition}{Proposition}
\newtheorem{theorem}{Theorem}
\newtheorem{lemma}{Lemma}
\newtheorem{corollary}{Corollary}

\usepackage{soul}

\begin{document}
\title{Achieving Covert Communication With A Probabilistic Jamming Strategy}
\author{
  Xun~Chen,
  Fujun~Gao,
  Min~Qiu,~\IEEEmembership{Member,~IEEE,}
  Jia~Zhang,~\IEEEmembership{Member,~IEEE,}\\
  Feng~Shu,~\IEEEmembership{Member,~IEEE,}
  and~Shihao~Yan,~\IEEEmembership{Senior Member,~IEEE}

%\thanks{This work is supported in part by Hainan Provincial Natural Science Foundation of China under Grant 621MS016, Scientific Research Foundation of Hainan University under Grant KYQD(ZR)-21014, Hunan Provincial Education Science "13th Five-Year Plan" Project under Grant XJK19BXX001, Hunan Provincial Education Science "13th Five-Year Plan" Project under Grant XJK20BXX002, Hunan Provincial Education Science "14th Five-Year Plan" Project under Grant XJK23AJD021, National Natural Science Foundation of China under Grant 61977062, National Natural Science Foundation of China under Grant 62177046, Natural Science Foundation of Hunan Province under Grant 2021JJ30866, Ministry of Education Science and Technology Development Center New Generation Information Technology Innovation Project under Grant cxy0269. \textit{(Corresponding author: .)}}
\thanks{Part of this work has been accepted by the 2023 IEEE Global Communications Conference (GLOBECOM) \cite{conf_covert2023}.}
\thanks{X.~Chen, F.~Gao and F.~Shu are with the School of Information and Communication Engineering, Hainan University, Haikou 570228, China. M.~Qiu is with the School of Electrical Engineering and Telecommunications, UNSW, NSW 2052, Australia. J.~Zhang is with the College of Information Science and Engineering, Shandong Normal University, Jinan, China. S.~Yan is with the School of Science and Security Research Institute, Edith Cowan University, Perth, WA 6027, Australia. (e-mails: \{chenxun, gaofj\}@hainanu.edu.cn, min.qiu@unsw.edu.au, zhangjia@sdnu.edu.cn, shufeng0101@163.com, s.yan@ecu.edu.au.)}
}
\maketitle

\begin{abstract}
In this work, we consider a covert communication scenario, where a transmitter Alice communicates to a receiver Bob with the aid of a probabilistic and uninformed jammer against an adversary warden's detection. The transmission status and power of the jammer are random and follow some priori probabilities. We first analyze the warden's detection performance as a function of the jammer's transmission probability, transmit power distribution, and Alice's transmit power. We then maximize the covert throughput from Alice to Bob subject to a covertness constraint, by designing the covert communication strategies from three different perspectives: Alice's perspective, the jammer's perspective, and the global perspective. Our analysis reveals that the minimum jamming power should not always be zero in the probabilistic jamming strategy, which is different from that in the continuous jamming strategy presented in the literature. In addition, we prove that the minimum jamming power should be the same as Alice's covert transmit power, depending on the covertness and average jamming power constraints. Furthermore, our results show that the probabilistic jamming can outperform the continuous jamming in terms of achieving a higher covert throughput under the same covertness and average jamming power constraints.
\end{abstract}

\section{Introduction}
Covert communication, also named low probability of detection (LPD) communication, is to hide the very existence of transmissions with proven performance. This can mitigate the threat of discovering the presence of a user or communication to achieve a high level of security and privacy, which is especially suitable for ensuring user privacy and information security in wireless networks \cite{8883125}. In recent years, with the rapid development and wide application of wireless communication technologies, an increasing amount of research has focused on wireless covert communication \cite{8883125}.

A pioneering work for covert communication over the additive white Gaussian noise (AWGN) channel was conducted in \cite{6584948}, which has proved that an arbitrarily LPD is possible if the transmitter sends at most $\mathcal{O}(\sqrt{n})$ bits over $n$ channel uses to the receiver. This result, known as the square-root law, has been shown to hold true for various channel models, such as the binary symmetric channel without a secret key \cite{6620765}, discrete memoryless channel \cite{7447769}, multi-access channel \cite{7541695}, multiple-input multiple-output AWGN channel \cite{8228657}, and the relay channel \cite{8736032,9438645,9363936}.

To improve the performance of covert communication, several works exploited the uncertainty in the adversary's observations in terms of noise power \cite{7805182}, communication channels \cite{8471218}, artificial noise (AN) \cite{7964713,8519751}, and transmission time \cite{9875023}. Notably, \cite{7964713} studied covert communication in the presence of a uninformed jammer that generates AN with randomized power and showed that a positive covert rate is achievable. Meanwhile, \cite{8519751} employed a full-duplex receiver to generate AN with varying power to enhance covert communication performance. In \cite{9007017}, the optimal power adaptation scheme of a legitimate transmitter was developed in term of minimizing the outage probability subject to covertness and average power constraints. When equipped with multiple antennas, the optimal strategy of the jamming is to perform beamforming towards a single direction with all the available power \cite{8849607}. Besides, covert communication with a finite blocklength was studied in multiple works (e.g., \cite{8379465,8878022,9444350}).

It is worth emphasizing that jamming signals transmitted by a friendly jammer can also become interference to the legitimate receiver. To reduce the detrimental effects of jamming on the legitimate receiver while enhancing covertness at the same time, \cite{10039177} and \cite{10058980} considered using a probabilistic jammer which emits the jamming signals with a certain probability. This is in sharp contrast to conventional continuous jamming that always has a transmission probability of one. Specifically, \cite{10039177} adopted a probabilistic jamming scheme with fixed power to aid covert communication in the finite blocklength regime, where the average effective covert throughput was maximized by optimizing the transmit power and blocklength. Meanwhile, \cite{10058980} demonstrated the superiority of covert communication aided by a probabilistic jammer with varying jamming power over that aided by a conventional continuous jammer in terms of achieving a higher energy efficiency.

A probabilistic jammer decides whether to transmit AN with a prior probability, while a continuous jammer can be regarded as a special case of the probabilistic jammer with one as the prior probability. Thus, a probabilistic jammer represents a new generalized jamming strategy. The power of a probabilistic jammer can follow a specific distribution, e.g., Bernoulli distribution \cite{10039177} or a mixed distribution of a Bernoulli distribution and a uniform distribution over interval $[0, P_{\max}]$ \cite{10058980}. Taking the AWGN channel as an example, if the transmission probability of AN is very low, the adversary warden will be able to determine the legitimate user's transmission status with a very low probability of error. On the other hand, a high transmission probability of AN will bring a high probability of having interference to the receiver and also degrade the legitimate communication performance. Therefore, in the probabilistic jamming strategy, the prior transmission probability of AN needs to be optimized to achieve the best balance between enhancing covertness and communication. Considering the limitations (e.g., probabilistic jamming with fixed power, minimum AN power set as zero) of the pioneering works \cite{10039177} and \cite{10058980} on the probabilistic jamming strategy, its benefits in the context of covert communications have not been fully revealed, which mainly motivates this work.

%However, neither the probabilistic jamming models nor the optimization problems employed in \cite{10039177} and \cite{10058980} are generic enough.

%The fixed power can be regarded as following a special continuous distribution in the sense that the power of a probabilistic jammer can be regarded as following a mixed distribution of a Bernoulli discrete distribution and a continuous distribution.

Motivated by the above promising potential benefits, we aim to further investigate the performance of covert communication based on the probabilistic jammer strategy. For generality, we consider that the transmission status of the jammer follows a Bernoulli distribution while the AN power follows a uniform distribution over the interval of the minimum and maximum jamming power $[P_{\min}, P_{\max}]$. Meanwhile, we consider two practical constraints, i.e., the covertness constraint and the average jamming power constraint. The main contributions of this work are summarized as follows:
\begin{itemize}
  \item We analyze the detection performance of the adversary warden and the communication performance of covert communication over the AWGN channel, where a radiometer is used as the detector. Specifically, we analytically derive the optimal detection threshold, warden's minimum detection error probability, covert throughput, covertness constraint, and design the transmission scheme of covert communication.
  \item Under the average jamming power constraint, we maximize the covert throughput from Alice's perspective, the jammer's perspective, and also from the jointly global point of view of both Alice and the jammer. For the maximization problems, we derive the feasible ranges of values for the optimizable parameters and derive the optimal designs of covert communication from the aforementioned three different perspectives. In addition, we show that from the jammer's perspective, the optimal design of covert communication is not unique.
  \item Both our analytical and numerical results demonstrate the superiority of the probabilistic jamming strategy over the continuous jamming strategy in terms of  achieving a higher covert throughput. Multiple extra insights on the probabilistic jamming strategy in covert communications have been provided. For example, it is revealed that, in the probabilistic jamming strategy, the jammer's minimum AN transmit power is not always zero but the same as Alice's covert transmit power, which depends on the required covertness level and the available average jamming power.
\end{itemize}

\section{System Model}
\subsection{Communication Scenario and Assumptions}
\begin{figure}[ht]
  \centering
  \vspace{-4pt}
  \setlength{\abovecaptionskip}{-2pt}
  \includegraphics[width=0.66\linewidth]{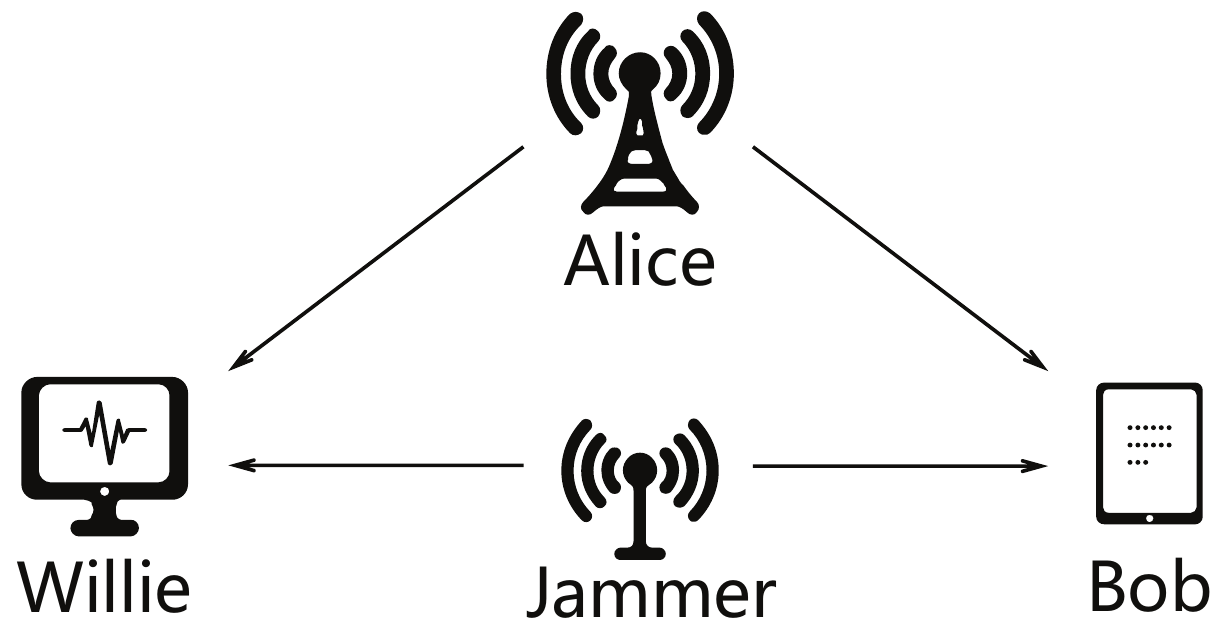}
  \caption{Covert communication model with a friendly jammer.}\label{fig:system_model}
  %\vspace{-4pt}
\end{figure}

We consider a covert communication network consisting of four single-antenna nodes, a transmitter (Alice), a receiver (Bob), a warden (Willie), and a friendly jammer (Jammer). The system model is depicted in Fig. \ref{fig:system_model}. Alice aims to transmit messages to Bob covertly under the surveillance of Willie. Willie tries to determine whether Alice transmits messages, by means of detecting any transmission from Alice. Jammer generates AN to deteriorate the detection performance of Willie for assisting the covert transmission from Alice to Bob.
\par
In order to focus on the impact of Jammer's signaling strategy on covert communication, we consider the AWGN channel model in this work. We use ${s_k}$ and $p_k$ to denote the transmission status of the transmitter $k\in\{a,j\}$ and its prior transmission probability, respectively, where $k=a$ represents Alice and $k=j$ represents Jammer with ${s_j} \sim \text{Bernoulli}({p_j})$. Alice and Jammer has no knowledge on the transmission status of each other. Thus, a reasonable assumption is that ${s_a}$ and ${s_j}$ are independent of each other. This means that Alice and Jammer do not cooperate with each other.

We assume that each transmitter adopts complex Gaussian signaling \cite{8379465}. That is, $\mathbf{x}_k[i] \sim \mathcal{CN}(0, {P_k})$, where $P_k$ is the transmit power of the transmitter $k$ and $i=1,...,N$ represents the symbol index in one time slot and $N$ is the length of the symbol block. We also assume that the symbol block length is very large such that $N \rightarrow + \infty$. The signal received at receiver $l$ for the $i$-th symbol period is given by
\begin{equation}\label{eqn:yk}
  %{y_k[i]} = {h_{ak}}{s_a}{x_a[i]}+ {h_{jk}}{s_j}{x_j[i]}+{r_k[i]},
  {\mathbf{y}_l[i]} = {s_a}{\mathbf{x}_a[i]}+{s_j}{\mathbf{x}_j[i]}+{\mathbf{r}_l[i]},
\end{equation}
where $l$ can be $w$ or $b$, representing Willie or Bob respectively, and ${\mathbf{r}_l}[i] \sim \mathcal{CN}(0, {\sigma_l^2})$ is the  AWGN noise at receiver $l$ with variance $\sigma_l^2$. We note that ${\mathbf{x}_a}$ and ${\mathbf{x}_j}$ are independent of each other. We assume that $P_a$ is fixed and $P_j$ is uniformly distributed over different time slots in the interval $[P_{\min},P_{\max}]$, i.e., $P_j \sim \mathcal{U}(P_{\min},P_{\max})$. Meanwhile, Willie is aware of the value of $P_a$ and the distribution of $P_j$. For given $s_a$ and $s_j$, ${\mathbf{y}_l[i]}$ follows a complex Gaussian distribution, i.e.,
 \begin{equation}
  {\mathbf{y}_l[i]} \sim \mathcal{CN}(0,{s_a}{P_a}+{s_j}{P_j}+{\sigma_l^2}),\ l\in\{b,w\}.\label{eqn:yk_distribution}
\end{equation}

\subsection{Willie's Detection Scheme}
Willie aims to determine whether Alice is transmitting in a certain time slot based on the received signal $\mathbf{y}_w$, i.e. ${s_a} = 0$ or ${s_a} = 1$. Thus, Willie faces a binary hypothesis testing problem, which is given by
\begin{equation}
  \begin{cases}
     {\mathcal{H}_0}:&{\mathbf{y}_w[i]} = {s_j}{\mathbf{x}_j[i]}+{\mathbf{r}_w[i]},\\
     {\mathcal{H}_1}:&{\mathbf{y}_w[i]} = {\mathbf{x}_a[i]}+{s_j}{\mathbf{x}_j[i]}+{\mathbf{r}_w[i]},
     \label{eqn:hypothesis_testing_1}
  \end{cases}
\end{equation}
where ${\mathcal{H}_0}$ is the null hypotheses and denotes that Alice has not transmitted, and ${\mathcal{H}_1}$ is the alternative hypothesis and denotes that Alice has transmitted.
\par
We assume that Willie adopts a radiometer \cite{7964713,8519751} to detect the covert communication from Alice to Bob, due to its low complexity and ease of implementation. Thus, Willie employs its average receive power ${P_w}$ as the test statistic to conduct a threshold test. In each time slot, ${P_w}$ is given by
\begin{equation}
  {P_w} = \frac{1}{N}\sum\limits_{i = 1}^{N}{|{\mathbf{y}_w[i]}|^2}.
\end{equation}
As per (\ref{eqn:yk_distribution}), ${\mathbf{y}_w[i]}$ follows a complex Gaussian distribution and thus ${P_w}$ is a chi-squared random variable with $2N$ degrees of freedom multiplied by a constant, i.e.,
\begin{equation}
  {P_w}=\frac{{s_a}{P_a}+{s_j}{P_j}+{\sigma_w^2}}{2N}{\chi^2}(2N),
\end{equation}
where ${\chi^2}(2N)$ represents a chi-squared random variable with $2N$ degrees of freedom. According to the Strong Law of Large Numbers, as $N\rightarrow\infty$, we have
\begin{equation}
  {P_w}\xrightarrow{P}{s_a}{P_a} + {s_j}{P_j} + {\sigma_w^2}.
\end{equation}
Then, the problem of binary hypothesis testing in (\ref{eqn:hypothesis_testing_1}) can be rewritten as
\begin{equation}
  \begin{cases}
    {\mathcal{H}_0}:&{P_w} = {s_j}{P_j} + {\sigma_w^2},\\
    {\mathcal{H}_1}:&{P_w} = {P_a} + {s_j}{P_j} + {\sigma_w^2},
     \label{eqn:hypothesis_testing_2}
  \end{cases}
\end{equation}
which is a binary composite hypothesis testing. Under both hypothesis, ${P_w}$ is jointly determined by ${s_j}$ and ${P_j}$, and thus it leads to a mixed discrete-continuous distribution. The probability density functions (pdfs) of $P_w$ under hypothesis ${\mathcal{H}_0}$ and ${\mathcal{H}_1}$ are given, respectively, as
\begin{subequations}
\begin{align}
  {f_{{P_w}}}\left( x|{\mathcal{H}_0} \right) =& {q_j}\delta \left( {x - \sigma _w^2} \right) + {p_j}{f_{{P_j}}}\left( {x - \sigma _w^2} \right),\label{eqn:cpdf_H0}\\
  {f_{{P_w}}}\left( x|{\mathcal{H}_1} \right) =& {q_j}\delta \left( {x - {P_a} - \sigma _w^2} \right)\notag\\
   &+ {p_j}{f_{{P_j}}}\left( {x - {P_a} - \sigma _w^2} \right),\label{eqn:cpdf_H1}
\end{align}
\end{subequations}
where we define $q_j\triangleq1-p_j$, $\delta \left( x \right)$ is the unit impulse function and $f_{P_j}\left( x \right)$ is the generalized pdf of $P_j$ given as
\begin{equation}
  {f_{P_j}}\left( x \right) = \frac{u\left ( {x - {P_{\min }}} \right ) - u\left ( {x - {P_{\max }}} \right )}{P_L},\label{eqn:pdf_Pj}
\end{equation}
where $u \left( x \right)$ is the unit step function and $P_L\triangleq P_{\max } - P_{\min }$.

In the radiometer detector, Willie's decision rule is given by
\begin{equation}
  P_w \overset{\mathcal{D}_1}{\underset{\mathcal{D}_0}{\gtreqless }} \gamma,\label{eqn:decision_rule}
\end{equation}
where $\mathcal{D}_0$ and $\mathcal{D}_1$ are the binary decisions that infer $\mathcal{H}_0$ and $\mathcal{H}_1$, respectively, and $\gamma$ is the detection threshold.
\par
In this work, the total detection error probability is adopted as the metric on Willie's detection performance \cite{6584948}, which is defined as
\begin{equation}
  \xi \triangleq {\mathbb{P}_{FA}}+{\mathbb{P}_{MD}},\label{eqn:TDEP}
\end{equation}
where $\mathbb{P}_{FA}={\Pr}(\mathcal{D}_1|\mathcal{H}_0)$ denotes the false alarm probability and $\mathbb{P}_{MD}={\Pr}(\mathcal{D}_0|\mathcal{H}_1)$ denotes the miss detection probability. Thus, ${\mathbb{P}_{FA}}$ and ${\mathbb{P}_{MD}}$ are equally important for Willie. Willie's ultimate goal is to detect the presence of Alice's transmission with the minimum total detection error probability $\xi^\ast$. To this end, Willie needs to obtain the optimal threshold $\gamma^\ast$ to minimize $\xi$. Therefore, the general covertness constraint is given by $\xi^\ast \ge 1-\epsilon$ for any $\epsilon>0$, where $\epsilon$ denotes the predetermined minimum covertness level.

\subsection{Transmission from Alice to Bob}
%Covert communication needs to satisfy $\xi^\ast \ge 1-\epsilon$ for any $\epsilon\in(0,1)$, where $\epsilon$ denotes the predetermined minimum covertness level. When both Willie and Alice adopt $\xi$ as the metric of detection error probability, Willie expects $\xi$ to be minimized while Alice expects it to be maximized. In other words, $\xi^\ast$ represents the best-case performance from Willie's perspective, but worst-case from Alice's, conversely. Hence, $\xi^\ast$ serves as a metric of covertness.
We assume that Alice transmits its messages with a fixed-rate to Bob. The covert throughput is employed to evaluate the communication performance from Alice to Bob \cite{8654724}, which is defined as
\begin{equation}
  \Omega \triangleq R \left(1-\lambda  \right),\label{eqn:covert_throughput}
\end{equation}
where $R$ denotes the transmission rate of Alice, and $\lambda $ denotes the transmission outage probability. A transmission outage occurs when
$C < R$, where $C$ is the channel capacity, and thus $\lambda={\Pr}\left(C < R\right)$. The channel capacity between Alice and Bob varies with the AN power. Thus the outage is caused by the AN, since the AN power is unknown to Bob. As per (\ref{eqn:yk}), we have $C=\log_2(1+\frac{P_a}{s_jP_j+\sigma^2_b})$.

\section{Performance Analysis on Covert Communication}
In this section, we analyze the performance of covert communication, where Willie's optimal detection threshold $\gamma^\ast$ is derived simultaneously when analyzing the minimum total detection error probability $\xi^\ast$.
\subsection{Covertness Analysis}
We first derive the expression of $\xi^\ast$, based on which we tackle the covertness constraint $\xi^\ast \ge 1-\epsilon$.
For the convenience of deriving $\xi^\ast$ and $\gamma^\ast$, we define
\begin{equation}
  \eta \triangleq \int_{\gamma - {P_a}-\delta} ^{\gamma-\delta } {{f_{{P_w}}} ({x|\mathcal{H}_0})} dx,
\end{equation}
where the notation $-\delta$ for an arbitrarily small $\delta>0$ represents the left limit and $\eta^\ast \triangleq \mathop {\max}\limits_{\gamma} \eta$. We then have the following lemma.
\begin{lemma}\label{lemma:min_xi}
 The minimum detection error probability is given by $\xi^\ast = 1 - \eta^\ast$.
\end{lemma}
\begin{IEEEproof}
First, we note that ${f_{{P_w}}}\left( x|{\mathcal{H}_1} \right)$ can be obtained by shifting ${f_{{P_w}}}\left( x|{\mathcal{H}_0} \right)$ to the right by ${P_a}$, i.e.,
\begin{equation}
  {f_{{P_w}}}\left( x|{\mathcal{H}_1} \right) = {f_{{P_w}}}\left( x - {P_a}|{\mathcal{H}_0} \right). \label{eqn:pdf_shift}
\end{equation}
Hence, we have
  \begin{equation}\label{eq:pw_H1}
    \begin{split}
     {\Pr}(P_w < \gamma|\mathcal{H}_1) =& \int_{-\infty } ^{\gamma -\delta} {{f_{{P_w}}} ({x|\mathcal{H}_1})} dx\\
    =&\int_{-\infty } ^{\gamma -\delta} {{f_{{P_w}}} ({x - {P_a}|\mathcal{H}_0})} dx\\
    =&\int_{-\infty } ^{\gamma - {P_a}-\delta} {{f_{{P_w}}} ({x|\mathcal{H}_0})} dx\\
    =&{\Pr}(P_w < \gamma - {P_a}|\mathcal{H}_0).
    \end{split}
  \end{equation}
Substituting \eqref{eqn:pdf_shift} into \eqref{eqn:TDEP}, we have
  \begin{equation}
    \begin{split}
     \xi =& {\Pr}(P_w \ge \gamma|\mathcal{H}_0) + {\Pr}(P_w < \gamma|\mathcal{H}_1)\\
     =&{\Pr}(P_w \ge \gamma|\mathcal{H}_0) + {\Pr}(P_w < \gamma - {P_a}|\mathcal{H}_0)\\
     =&1-{\Pr}(\gamma - {P_a} \le P_w < \gamma |\mathcal{H}_0)\\
     =&1-\int_{\gamma - {P_a}-\delta} ^{\gamma -\delta } {{f_{{P_w}}} ({x|\mathcal{H}_0})} dx\\
     =&1-\eta.
    \end{split}
  \end{equation}
Then, the minimum detection error probability is
  \begin{equation}
    \xi^\ast=\mathop{\min}\limits_{\gamma}\xi=1-\mathop{\max}\limits_{\gamma}\eta=1-\eta^\ast.
  \end{equation}
This completes the proof.
\end{IEEEproof}
\par
%\textit{Lemma \ref{lemma:min_xi}} mainly makes use of the fact that $f_{P_j}\left( x \right)$ is constant in the distribution interval, and the translation property between the likelihood functions ${f_{P_w}}( x|{\mathcal{H}_0} )$ and ${f_{P_w}}( x|{\mathcal{H}_1} )$.

Following \textit{Lemma \ref{lemma:min_xi}}, we derive $\xi^\ast$ and $\gamma^\ast$ in the following theorem.

\begin{theorem}\label{theorem:min_xi}
  The minimum total detection error probability $\xi^\ast$, and the optimal threshold $\gamma^\ast$, are given in Table \ref{tab:opt_par} at the top of next page.

  \begin{table*}[htb]
    \begin{center}
    \begin{threeparttable}
    %\vspace{-4pt}
      \caption{Expressions for $\gamma^\ast$ and $\xi^\ast$\label{tab:opt_par}}
      \renewcommand\arraystretch{1.8}
      \begin{tabular}{llll}
        \toprule
        Condition 1 & Condition 2 & $\gamma^\ast$ & $\xi^\ast$\\
        \hline
        \multirow{2}{*}{${P_a} \ge {P_{\max }}$} & ${p_j} < 1$ & $[\sigma _w^2 + {P_{\max }},\sigma _w^2 + {P_a}]$ & \multirow{2}{*}{0}\\
        \cline{2-3}
        & ${p_j} = 1$& $[\sigma_w^2+ P_{\max }, \sigma_w^2+ P_{\min }+P_a]$& \\
        \hline
        \multirow{2}{*}{${P_a} \le \min ({P_{\min }},{P_L})$} & ${p_j} < \frac{{{P_L}}}{{{P_L} + {P_a}}}$ & $(\sigma _w^2,\sigma _w^2 + {P_a}]$ & $p_j$\\
        \cline{2-4}
        &${p_j} > \frac{{{P_L}}}{{{P_L} + {P_a}}}$ & $[\sigma _w^2 + {P_{\min }} + {P_a},\sigma _w^2 + {P_{\max }}]$ & $1 - {p_j}\frac{{{P_a}}}{{{P_L}}}$\\
        \hline
        \multirow{2}{*}{${P_{\min }} < {P_a} \le {P_L}$} & ${p_j} < \frac{{{P_L}}}{{{P_{\max }}}}$ & $\sigma _w^2 + {P_a}$ & ${p_j}\frac{{{P_{\max }} - {P_a}}}{{{P_L}}}$\\
        \cline{2-4}
        & ${p_j} > \frac{{{P_L}}}{{{P_{\max }}}}$ & $[\sigma _w^2 + {P_{\min }} + {P_a},\sigma _w^2 + {P_{\max }}]$ & $1 - {p_j}\frac{{{P_a}}}{{{P_L}}}$\\
        \hline
        \multirow{2}{*}{${P_L} < {P_a} \le {P_{\min }}$} & ${p_j} < \frac{1}{2}$ & $(\sigma _w^2,\sigma _w^2 + {P_a}]$ & ${p_j}$\\
        \cline{2-4}
        & ${p_j} > \frac{1}{2}$ & $[\sigma _w^2 + {P_{\max }},\sigma _w^2 + {P_{\min }} + {P_a}]$ & ${q_j}$\\
        \hline
        \multirow{2}{*}{$\max ({P_{\min }},{P_L}) < {P_a} < {P_{\max }}$} & ${p_j} < \frac{{{P_L}}}{{{P_L} + {P_{\max }} - {P_a}}}$ & $\sigma _w^2 + {P_a}$ & ${p_j}\frac{{{P_{\max }} - {P_a}}}{{{P_L}}}$\\
        \cline{2-4}
        & ${p_j} > \frac{{{P_L}}}{{{P_L} + {P_{\max }} - {P_a}}}$ & $[\sigma _w^2 + {P_{\max }},\sigma _w^2 + {P_{\min }} + {P_a}]$ & ${q_j}$\\
        \bottomrule
      \end{tabular}
      \begin{tablenotes}
        \item [1] ${P_L}\triangleq {P_{\max }}-{P_{\min }}$.
        \item [2] When ${P_a} < {P_{\max }}$, condition 2 does not list the case where $=$ holds. When the expressions of $\xi^\ast$ in the two cases of $<$ and $>$ are equal, $\gamma^\ast$ is the union of the expressions in the two cases.
      \end{tablenotes}

    %\vspace{-10pt}
  \end{threeparttable}
  \end{center}
  \end{table*}
\end{theorem}
\begin{IEEEproof}
  The proof is provided in Appendix \ref{sec:appA}.
\end{IEEEproof}

As per Table \ref{tab:opt_par}, when $P_a \ge P_{\max}$, Willie achieves zero minimum detection error probability. Hence, $P_a < P_{\max}$ is the precondition for achieving any degree of covertness. Furthermore, we present the following corollary regarding the covertness constraint.
\begin{corollary}\label{corollary:covertness_constraint}
The covertness constraint $\xi^\ast \ge 1-\epsilon$ for $\epsilon \in \left(0,\frac{1}{2}\right)$ is equivalent to the following constraints on Jammer's transmission probability $p_j$, the minimum AN power $P_{\min}$, the maximum AN power $P_{\max}$, and Alice's transmit power $P_a$.
  \begin{equation}\label{eqn:covertness_constraints}
    \begin{cases}
      {p_j} \ge 1 - \epsilon,\\
      {P_{\max }} - {P_{\min }} \ge \frac{{{p_j}}}{\epsilon}{P_a},\\
      {\left( {\frac{{{p_j}}}{{1 - \epsilon}} - 1} \right){P_{\max }} + {P_{\min }} \ge \frac{{{p_j}}}{{1 - \epsilon}}{P_a}}.
    \end{cases}
  \end{equation}
\end{corollary}

\begin{IEEEproof}
For $\epsilon \in \left(0,\frac{1}{2}\right)$, we have $\xi^\ast \ge 1-\epsilon > \frac{1}{2}$. Thus, $\xi^\ast > \frac{1}{2}$ is necessary. As per Table \ref{tab:opt_par}, we analyze the covertness constraint in the following five cases.
  \begin{enumerate}
    \item Case ${P_a} \ge {P_{\max }}$: According to Table \ref{tab:opt_par}, $\xi^\ast = 0$ always holds, thus the covertness constraint $\xi^\ast \ge 1-\epsilon$ cannot be satisfied.
    \item Case ${P_L} < {P_a} \le {P_{\min }}$: If $p_j<\frac{1}{2}$, then $\xi^\ast = p_j <\frac{1}{2}$; otherwise $\xi^\ast = 1- p_j \le \frac{1}{2}$. Thus $\xi^\ast = \min\left(p_j,1-p_j\right) \le \frac{1}{2} < {1 - \epsilon}$, which does not satisfy the covertness constraint.
    \item Case $\max ({P_{\min }},{P_L}) < {P_a} < {P_{\max }}$: $\xi^\ast$ achieves the maximum value of $\frac{{P_{\max }} - {P_a}}{{P_L} +{P_{\max }} - {P_a}}$, when ${p_j} = \frac{{{P_L}}}{{{P_L} + {P_{\max }} - {P_a}}}$. Since $P_L > {P_{\max }} - {P_a}$, we have $\xi^\ast \le \frac{{P_{\max }} - {P_a}}{{P_L} +{P_{\max }} - {P_a}}  < \frac{1}{2} < {1 - \epsilon}$. Again, the covertness constraint cannot be satisfied.
    \item Case ${P_{\min }} < {P_a} \le {P_L}$: When $p_j=\frac{P_L}{P_{\max}}$, $\xi^\ast$ achieves the maximum value of $1-\frac{P_a}{P_{\max}}$ with respect to (w.r.t.) $p_j$. To guarantee $\xi^\ast \ge 1-\epsilon$, the followings should hold.
    \begin{equation}
      \begin{cases}
        1-\frac{P_a}{P_{\max}} \ge 1-\epsilon,\\
        {p_j}\frac{P_{\max} - {P_a}}{P_L} \!\ge\! 1-\epsilon,\\
        1 - {p_j}\frac{P_a}{P_L} \ge 1-\epsilon,
      \end{cases}
     \Rightarrow  \begin{cases}
      {p_j} > 1 - {\epsilon},\\
      \frac{P_L}{P_a} \ge \frac{p_j}{\epsilon},\\
      \frac{{P_{\min }} - {P_a}}{{P_{\max }} - {P_a}} \ge 1 - \frac{p_j}{1 - \epsilon}.
      \end{cases}
    \end{equation}
    \item Case ${P_a} \le \min ({P_{\min }},{P_L})$: When $p_j=\frac{{P_L}}{{P_L} + {P_a}}$, $\xi^\ast$ achieves the maximum value of $\frac{{{P_L}}}{{{P_L} + {P_a}}}$ w.r.t. $p_j$. To guarantee $\xi^\ast \ge 1-\epsilon$, the followings should hold.
    \begin{align}
      \begin{cases}
        \frac{{{P_L}}}{{{P_L} + {P_a}}} \ge 1-\epsilon,\\
        p_j \ge 1-\epsilon,\\
        1 - {p_j}\frac{{{P_a}}}{{{P_L}}} \ge 1-\epsilon,
      \end{cases}
      \Rightarrow \begin{cases}
        p_j \ge 1-\epsilon,\\
        \frac{P_L}{P_a} \ge \frac{p_j}{\epsilon}.
      \end{cases}
    \end{align}
  \end{enumerate}

From above, we see that the covertness constraint $\xi^\ast \ge 1-\epsilon$ can only be satisfied in cases 4) and 5). By combing the results of the last two cases above, we obtain \eqref{eqn:covertness_constraints}.
\end{IEEEproof}
%\begin{remark}
%  The minimum covertness requirement determines the minimum feasible transmission probability of AN, i.e., ${p_j} \ge 1 - {\epsilon}$. To satisfy the requirement $\epsilon \in \left(0,\frac{1}{2}\right)$, two case are involved, i.e., ${P_{\min }} < {P_a} \le {P_L}$ and ${P_a} \le \min ({P_{\min }},{P_L})$, which are combined into Equation (\ref{eqn:covertness_constraints}). It is worth noting that ${p_j} = 1 - {\epsilon}$ holds only if ${P_a} \le \min ({P_{\min }},{P_L})$.
%\end{remark}
\textit{Corollary \ref{corollary:covertness_constraint}} provides the necessary and sufficient conditions for covert communication to satisfy the covertness requirement $\epsilon \in \left(0,\frac{1}{2}\right)$, which will be used for solving the optimization problem in Section \ref{sec:opt}.

\subsection{Covert Communication Scheme Design}
In this subsection, we first derive the expression for the covert throughput $\Omega$. Since the transmission rate $R$ is independent of covertness constraint, we can maximize $\Omega$ by designing $R$, which leads to the optimal value of $R$.

When Alice transmits, Jammer is either active or silent. Applying the Law of Total Probability, the transmission outage probability can be written as
\begin{equation}
\begin{split}
  %\lambda =&\Pr \left\{ {C < R} \right\}\\
  \lambda=&{q_j}\Pr \left\{ {C < R|s_j=0} \right\} + {p_j}\Pr \left\{ {C < R|s_j=1} \right\}\\
  %=&{q_j}\Pr \left\{ {{{\log }_2}\left( {1 + \frac{{{P_a}}}{{\sigma _b^2}}} \right) < R} \right\} \\
  %&+ {p_j}\Pr \left\{ {{{\log }_2}\left( {1 + \frac{{{P_a}}}{{{P_j} + \sigma _b^2}}} \right) < R} \right\}\\
  =&{q_j}\Pr \left\{ {P_r} < 0 \right\} + {p_j}\Pr \left\{ {P_j} > {P_r} \right\}\\
  =& \left\{
  \begin{aligned}
      &0,&{P_r} \ge {P_{\max}}, \\
      &{p_j}\frac{{{P_{\max }} - {P_r}}}{{{P_{\max }} - {P_{\min }}}},&{P_{\min}} \le {P_r} \le {P_{\max}},\\
      &{p_j},&0 \le {P_r} \le {P_{\min}},\\
      &1,&{P_r} < 0,
  \end{aligned}
  \right.\\
  =& \left\{
  \begin{aligned}
      &0,&R \le C_n, \\
      &{p_j}\frac{{{P_{\max }} - {P_r}}}{{{P_{\max }} - {P_{\min }}}},&C_n \le R \le C_j,\\
      &{p_j},& C_j \le R \le C_f,\\
      &1,&R > C_f,
  \end{aligned}
  \right.
\end{split}\label{eqn:outage_robability}
\end{equation}
where we define ${P_r} \triangleq \frac{{{P_a}}}{{{2^R} - 1}} - \sigma _b^2$, ${C_n} \triangleq {\log_2}\left( 1 + {\frac{P_a}{\sigma _b^2 +P_{\max }}} \right)$, ${C_j} \triangleq {\log_2}\left( 1 + {\frac{P_a}{\sigma _b^2 +P_{\min }}} \right)$, and ${C_f} \triangleq {\log_2}\left( 1 + {\frac{P_a}{\sigma _b^2}} \right)$. It is worth noting that ${C_n}$ is the minimum channel capacity for $s_j=1$, which represents the maximum $R$ that guarantees $\lambda=0$. In addition, ${C_j}$ is the maximum channel capacity for $s_j=1$, which represents the maximum $R$ that guarantees $\lambda<1$ for $p_j<1$. Meanwhile, ${C_f}$ is the maximum channel capacity of all the time, which represents the minimum $R$ that results in $\lambda=1$.

Combining (\ref{eqn:covert_throughput}) and (\ref{eqn:outage_robability}), we obtain the covert throughput given by
%\begin{lemma}\label{lemma:throughput}
  \begin{equation}
      \Omega = \left \{
        \begin{aligned}
          & R,& R \le C_n, \\
          & R  \left( 1 -{p_j}\frac{{{P_{\max }} - {P_r}}}{{{P_{\max }} - {P_{\min }}}} \right) , &C_n \le R \le C_j,\\
          & {q_j} R,&C_j \le R \le C_f,\\
          &0,&R > C_f.
        \end{aligned}
        \right.
        \label{eqn:omega}
      \end{equation}
%\end{lemma}
\par
  For given $\sigma _b^2$, ${P_a}$, ${P_{\min }}$ and ${P_{\max }}$, $\Omega$ is a piecewise and continuous function w.r.t. $R$. Thus, $R$ can be optimized to achieve the maximum covert throughput. Let $\Omega_f \triangleq \Omega|_{R=C_f}={q_j} C_f$ and $\Omega_n \triangleq \Omega|_{R=C_n}=C_n$, we have the following proposition.
\begin{proposition}\label{prop:omega}
  The optimal value of the transmission rate $R$ is either $C_n$ or $C_f$. This implies that the maximum value of the throughput $\Omega$ is either $\Omega_f$ or $\Omega_n$.
\end{proposition}
\begin{IEEEproof}
  As per (\ref{eqn:omega}), when $0<p_j<1$ and $P_{\min } > 0$, $\Omega$ is a piecewise and continuous function of $R\in[0,C_f]$. Moreover, $\Omega$ is a strictly monotonically increasing function of $R$ in both the intervals $[0,C_n]$ and $[C_j,C_f]$. In interval $[C_n,C_j]$, the second derivative of $\Omega$ w.r.t. $R$ is derived as
  \begin{equation}\label{eqn:sec_der}
    \Omega ^{''}(R)= - {p_j}\frac{P_a}{P_L}\frac{{2^R}\ln 2}{{\left( {{2^R} - 1} \right)}^3} \omega (R),
  \end{equation}
  where $\omega (R)\triangleq 2^{R + 1} - {2^R}R\ln 2 - R\ln 2 - 2$. The first and second derivatives of $\omega (R)$ w.r.t $R$ are $\omega ^{'}(R)=\left( {{2^R} - {2^R}R {\ln 2}  - 1} \right){\ln 2}$ and $\omega ^{''}(R)=-{2^R}R{\left( {\ln 2} \right)^3}$, respectively. For $R>0$, we have
\begin{equation}
  \begin{cases}
    \omega ^{'}(0)=0\\
    \omega ^{''}(R)<0
  \end{cases}
  \Rightarrow
    \omega ^{'}(R)<\omega ^{'}(0)=0,
\end{equation}
and
\begin{equation}\label{eqn:wR}
  \begin{cases}
    \omega (0)=0\\
    \omega ^{'}(R)<0
  \end{cases}
  \Rightarrow
  \omega (R) < \omega (0) = 0.
\end{equation}
Substituting (\ref{eqn:wR}) into (\ref{eqn:sec_der}), we have $\Omega ^{''}(R)>0$ for $R>0$. Hence, $\Omega$ is convex w.r.t $R$ in the interval $[C_n,C_j]$. Combining with the fact that $\Omega$ monotonically increases with $R$ for $R\in [0,C_n]$ and $R\in [C_j,C_f]$, we obtain the conclusion stated in \textit{Proposition \ref{prop:omega}}. For $p_j=1$ or $P_{\min } = 0$, the same conclusion still holds.
\end{IEEEproof}
\par
Following \textit{Proposition \ref{prop:omega}}, we note that the maximum covert throughput is achieved when the transmission rate of Alice is set to the minimum or the maximum channel capacity. When AN was transmitted continuously, the optimal transmission rate is the minimum channel capacity $C_n$, because setting the transmission rate to the maximum channel capacity $C_j$ will lead to the outage probability being one. Therefore, using a probabilistic jammer in covert communication gives another degree-of-freedom in designing the optimal transmission rate $R$. This is different from the scenario with a conventional continuous jammer.

\section{Optimization of Covert Throughput}\label{sec:opt}
In this section, we maximize the covert throughput $\Omega$ subject to a given covertness requirement $\epsilon$ and a given average jamming power constraint $P_m$. The optimization problem is formulated as
\begin{align}
  \text{(P1):} \quad & \mathop {\text{maximize  }}\limits_{R,P_a,P_{\min},P_{\max},p_j} \Omega &\notag\\
  &\text{s.t.} \quad
  \begin{aligned}[t]
      &\text{(S1):} \quad \xi^\ast \ge 1-\epsilon ,\\
      &\text{(S2):} \quad \frac{1}{2} p_j \left(P_{\min} + P_{\max}\right) \le P_m,
  \end{aligned}
  &\label{p1}
\end{align}
where (S2) represents Jammer's average power constraint and $P_m$ denotes the maximum average transmit power of Jammer.
\par
In the following three subsections, we solve the optimization problem (P1) from Jammer's perspective, Alice's perspective, and the global perspective, respectively. That is, we investigate the optimal design to maximize the covert throughput from the point view of Jammer, Alice, and both Jammer and Alice, respectively.

\subsection{Optimal Design at Jammer}
In this subsection, we aim at maximizing the covert throughput from Jammer's perspective by designing Jammer's optimal transmission probability $p_j^\ast$, the optimal minimum AN power ${P_{\min}^\ast}$, and the optimal maximum AN power ${P_{\max}^\ast}$, for given Alice's transmission rate $R$ and Alice's transmit power $P_a$. As per (\ref{eqn:covert_throughput}), to maximize the covert throughput for given $P_a$ and $R$ is to minimize the transmission outage probability $\lambda$. Thus, (P1) can be rewritten as
\begin{align}
  \text{(P1.1):} \quad & \mathop {\text{minimize  }}\limits_{P_{\min},P_{\max},p_j} \lambda &\notag\\
  &\text{s.t.} \quad \text{(S1)},\text{(S2)}.&
  \label{p1.1}
\end{align}
\par
First, we investigate the feasibility of the optimization problem (P1.1). Then, we analyze the feasible value ranges for $p_j$, $P_{\min}$, and $P_{\max}$.
\begin{lemma}\label{lemma:feasibility1}
  The feasible conditions for the optimization problem (P1.1) are given by
  \begin{equation}\label{eq:feasible}
    \begin{cases}
      {P_a} \le \frac{{2\epsilon}}{{1 - {\epsilon^2}}}{P_m},\\
      R \le {C_f}.
    \end{cases}
  \end{equation}
  For any ${P_a}$ and $R$ satisfying \eqref{eq:feasible}, the feasible value ranges for $p_j$, $P_{\min}$ and $P_{\max}$ are derived as
  \begin{equation}\label{eqn:pars_feasible}
    \begin{cases}
      1-\epsilon \le p_j \le p_{ju},\\
      \frac{1}{\epsilon}{P_a} \le {P_{\max }} \le \min\left( \frac{{2\left( {1 - \epsilon} \right){P_m} - p_j^2{P_a}}}{{2\left( {1 - \epsilon} \right){p_j} - p_j^2}}, \frac{2}{{{p_j}}}{P_m}\right),\\
      \max\left(0, \frac{{1 - \epsilon - {p_j}}}{{1 - \epsilon}}{P_{\max }} + \frac{{{p_j}}}{{1 - \epsilon}}{P_a}\right) \le {P_{\min }} \\
      \qquad \qquad \le \min\left({P_{\max }} - \frac{{{p_j}}}{\epsilon}{P_a},\frac{2}{{{p_j}}}{P_m} - {P_{\max }}\right),
    \end{cases}
  \end{equation}
  where
  \begin{equation}
    {p_{ju}} =
    \begin{cases}
      1,& {P_a} \le 2\epsilon{P_m},\\
      1 - \sqrt {1 - 2\epsilon\frac{{{P_m}}}{{{P_a}}}},& 2\epsilon{P_m} \le {P_a} \le \frac{{2\epsilon}}{{1 - {\epsilon^2}}}{P_m},
    \end{cases}
  \end{equation}
  is the maximum jamming transmission probability.
\end{lemma}
\begin{IEEEproof}
  The proof is provided in Appendix \ref{sec:appB}.
\end{IEEEproof}

  Following \textit{Lemma \ref{lemma:feasibility1}}, we note that, under Jammer's average power constraint, to satisfy the covertness constraint $\xi^\ast \ge 1-\epsilon$, Alice's transmission rate $R$ and transmit power $P_a$ must be small enough. Specifically, Alice's transmission rate $R$ must be small enough for non-zero covert throughput, while Alice's transmit power $P_a$ must be small enough to meet the covertness constraint.

Following \textit{Lemma \ref{lemma:feasibility1}}, we derive the following solution to the optimization problem (P1.1).
\begin{theorem}\label{sol:jammer}
Let ${P_r} \triangleq \frac{{{P_a}}}{{{2^R} - 1}} - \sigma _b^2$, ${C_\epsilon} \triangleq {\log_2}\left( 1 + {\frac{{\epsilon}{P_a}}{{\epsilon}{\sigma _b^2} + {P_a} }} \right)$, and ${C_a} \triangleq {\log_2}\left( 1 + {\frac{P_a}{\sigma _b^2 +{P_a}}} \right)$, from Jammer's perspective, the optimal design solutions can be given in the following four cases:
\begin{enumerate}
  \item Case $R<C_\epsilon$: %${P_r} > \frac{P_a}{\epsilon}$:
  \begin{equation}\label{eqn:sol_case1}
    \hspace{-8pt}
    \begin{cases}
      1-\epsilon \le p_j \le p_{ju},\\
      \frac{1}{\epsilon}{P_a}\! \le\! {P_{\max }} \!\le\! \min\!\left( \frac{{2\left( {1 - \epsilon} \right){P_m} - p_j^2{P_a}}}{{2\left( {1 - \epsilon} \right){p_j} - p_j^2}}\!, \!\frac{2}{{{p_j}}}{P_m}, \!{P_r}\right)\!,\\
    \max\left(0, \frac{{1 - \epsilon - {p_j}}}{{1 - \epsilon}}{P_{\max }} + \frac{{{p_j}}}{{1 - \epsilon}}{P_a}\right) \le {P_{\min }} \\
    \qquad \le \min\left({P_{\max }} - \frac{{{p_j}}}{\epsilon}{P_a},\frac{2}{{{p_j}}}{P_m} - {P_{\max }}\right).
  \end{cases}
  \end{equation}
  \item Case $C_\epsilon \le R<C_a$:%${P_a} < {P_r} \le \frac{P_a}{\epsilon}$:
  \begin{equation}\label{eqn:sol_case2}
    \begin{cases}
      1-\epsilon \le p_j \le p_{ju},\\
      {P_{\max }} = \frac{{P_a}}{\epsilon},\\
      {P_{\min }} = \frac{{1 - {p_j}}}{\epsilon}{P_a}.
    \end{cases}
  \end{equation}
  \item Case $R=C_a$:%${P_r} = {P_a}$:
  \begin{equation}\label{eqn:sol_case3}
    \hspace{-8pt}
    \begin{cases}
      1-\epsilon \le p_j \le p_{ju},\\
      \frac{1}{\epsilon}{P_a} \!\le\! {P_{\max }} \!\le\! \min\!\left( \frac{{2\left( {1 - \epsilon} \right){P_m} - p_j^2{P_a}}}{{2\left( {1 - \epsilon} \right){p_j} - p_j^2}}\!,\! \frac{{{p_j}}}{{{p_j} + \epsilon - 1}}\!{P_a}\right)\!,\\
      {P_{\min }}=\frac{{1 - \epsilon - {p_j}}}{{1 - \epsilon}}{P_{\max }} + \frac{{{p_j}}}{{1 - \epsilon}}{P_a}.
    \end{cases}
  \end{equation}
  \item Case $C_a < R \le C_f$:%$0 \le {P_r} < {P_a}$:
  \begin{equation}\label{eqn:sol_case4}
    \begin{cases}
      p_j = 1-\epsilon,\\
      {P_a} \le {P_{\min }} \le \frac{1}{{1 - \epsilon}}{P_m} - \frac{1 - \epsilon}{2\epsilon}{P_a},\\
      \frac{1-\epsilon}{\epsilon}{P_a} + {P_{\min }} \le {P_{\max }} \le \frac{2}{{1 - \epsilon}}{P_m} - {P_{\min }}.
    \end{cases}
  \end{equation}
\end{enumerate}
The corresponding maximum covert throughput $\Omega^\ast$ is given by
\begin{equation}
  \Omega^\ast=
  \begin{cases}
    R, &R \le C_\epsilon,\\%{P_r} \ge \frac{P_a}{\epsilon}\\
    \epsilon R\frac{{{P_r}}}{{{P_a}}}, &C_\epsilon \le R \le C_a,\\%{P_a} \le {P_r} \le \frac{P_a}{\epsilon}\\
    \epsilon R, &C_a \le R \le C_f.%0\le {P_r} \le {P_a}.
  \end{cases}
\end{equation}

% \begin{equation}
%   \begin{cases}
%     {P_r} > \frac{P_a}{\epsilon}:
%     \begin{cases}
%       1-\epsilon \le p_j \le p_{ju}\\
%       \frac{1}{\epsilon}{P_a} \le {P_{\max }} \le \min\left( \frac{{2\left( {1 - \epsilon} \right){P_m} - p_j^2{P_a}}}{{2\left( {1 - \epsilon} \right){p_j} - p_j^2}}, \frac{2}{{{p_j}}}{P_m}, {P_r}\right)\\
%     \max\left(0, \frac{{1 - \epsilon - {p_j}}}{{1 - \epsilon}}{P_{\max }} + \frac{{{p_j}}}{{1 - \epsilon}}{P_a}\right) \le {P_{\min }} \\
%     \qquad \le \min\left({P_{\max }} - \frac{{{p_j}}}{\epsilon}{P_a},\frac{2}{{{p_j}}}{P_m} - {P_{\max }}\right)
%   \end{cases}\\
%   {P_a} < {P_r} \le \frac{P_a}{\epsilon}:
%   \begin{cases}
%     1-\epsilon \le p_j \le p_{ju}\\
%     {P_{\max }} = \frac{{P_a}}{\epsilon}\\
%     {P_{\min }} = \frac{{1 - {p_j}}}{\epsilon}{P_a}
%   \end{cases}\\
%   {P_r} = {P_a}:
%   \begin{cases}
%     1-\epsilon \le p_j \le p_{ju}\\
%     \frac{1}{\epsilon}{P_a} \le {P_{\max }} \le \min\left( \frac{{2\left( {1 - \epsilon} \right){P_m} - p_j^2{P_a}}}{{2\left( {1 - \epsilon} \right){p_j} - p_j^2}}, \frac{{{p_j}}}{{{p_j} + \epsilon - 1}}{P_a}\right)\\
%     {P_{\min }}=\frac{{1 - \epsilon - {p_j}}}{{1 - \epsilon}}{P_{\max }} + \frac{{{p_j}}}{{1 - \epsilon}}{P_a}
%   \end{cases}\\
%   0 \le {P_r} < {P_a}:
%   \begin{cases}
%     p_j = 1-\epsilon\\
%     {P_a} \le {P_{\min }} \le \frac{1}{{1 - \epsilon}}{P_m} - \frac{1 - \epsilon}{2\epsilon}{P_a}\\
%     \frac{1-\epsilon}{\epsilon}{P_a} + {P_{\min }} \le {P_{\max }} \le \frac{2}{{1 - \epsilon}}{P_m} - {P_{\min }}
%   \end{cases}.
% \end{cases}
% \end{equation}
\end{theorem}
\begin{IEEEproof}
  The proof is provided in Appendix \ref{sec:appC}.
\end{IEEEproof}

Following \textit{Theorem \ref{sol:jammer}}, we note that, from Jammer's perspective, the optimal transmission probability $p_j^\ast$, the optimal minimum AN power ${P_{\min}^\ast}$, and the optimal maximum AN power ${P_{\max}^\ast}$ are not always unique. Hence, the expressions for their feasible value ranges are derived in different cases determined by the relationship between $P_a$ and $R$. It is worth noting that the proposed design with $p_j^\ast=1-\epsilon$, ${P_{\min}^\ast}=P_a$, and ${P_{\max}^\ast}=\frac{P_a}{\epsilon}$ is always optimal, which leads to the lowest average jamming power $\frac{1 - {\epsilon^2}}{{2\epsilon}}{P_a}$ and the lowest covertness level $1-\epsilon$. The reason is that when Alice's transmission parameters are given, maximizing covert throughput is equivalent to minimizing the outage probability. In addition, the jamming parameters of minimizing the outage probability are not always unique.

\subsection{Optimal Design at Alice}
In this subsection, we aim at maximizing the covert throughput from Alice's perspective. Specifically, we design Alice's transmission rate $R$ and her transmit power $P_a$, for given Jammer's transmission probability $p_j$, minimum AN power ${P_{\min}}$, and maximum AN power ${P_{\max}}$. Thus from Alice's point of view, the optimization problem (P1) can be reformulated as
\begin{align}
  \text{(P1.2):} \quad & \mathop {\text{minimize  }}\limits_{R,P_a} \Omega &\notag\\
  &\text{s.t.} \quad \text{(S1)},\text{(S2)}.&
  \label{p1.2}
\end{align}
We still solve problem (P1.2) in two steps. First, we analyze the feasibility of the problem and the feasible value ranges for the parameters of interest. Then, we derive the optimal values of $P_a$ and $R$.
\begin{lemma}\label{lemma:feasibility2}
  The feasible conditions for the optimization problem (P1.2) can be written as
  \begin{equation}
    \begin{cases}
      {p_j} > 1 - \epsilon,\\
      0 \le {P_{\min }} < \frac{1}{{{p_j}}}{P_m},\\
      {P_{\min }} < {P_{\max }} \le \frac{2}{{{p_j}}}{P_m} - {P_{\min }},
    \end{cases}
  \end{equation}
  and
  \begin{equation}
    \begin{cases}
      {p_j} = 1 - \epsilon,\\
      0 < {P_{\min }} < \frac{1}{{1 - \epsilon}}{P_m},\\
      {P_{\min }} < {P_{\max }} \le \frac{2}{{1 - \epsilon}}{P_m} - {P_{\min }},
    \end{cases}
  \end{equation}
  and the feasible value ranges of ${P_a}$ and $R$ can be given by
  \begin{equation}
    \begin{cases}
      {P_a} \le {P_{au}},\\
      R \le {C_f},
    \end{cases}
  \end{equation}
  where
  \begin{equation}\label{eq:Pau}
    {P_{au}}\!=\!
    \begin{cases}
      \!\left( {1 \!-\! \frac{{1 - \epsilon}}{{{p_j}}}} \right)\!{P_{\max }} \!+\! \frac{{1 - \epsilon}}{{{p_j}}}{P_{\min }}, &\frac{{{P_{\min }}}}{{{P_{\max }}}} \!\le \!1 \!-\! {p_j},\\
      \frac{\epsilon}{{{p_j}}}\!\left( {{P_{\max }} \!-\! {P_{\min }}} \right),&\frac{{{P_{\min }}}}{{{P_{\max }}}} \!\ge\! 1\! -\! {p_j},
    \end{cases}
  \end{equation}
  is Alice's maximum transmit power.
\end{lemma}
\begin{IEEEproof}
  Following \textit{Corollary \ref{corollary:covertness_constraint}}, constraints (S1) and (S2) can be rewritten together as
  \begin{equation}
    \begin{cases}
      {p_j} \ge 1 - \epsilon,\\
      {P_a} \le \frac{\epsilon}{{{p_j}}}\left( {{P_{\max }} - {P_{\min }}} \right),\\
      {P_a} \le \left( {1 - \frac{{1 - \epsilon}}{{{p_j}}}} \right){P_{\max }} + \frac{{1 - \epsilon}}{{{p_j}}}{P_{\min }}, \\
      {P_{\max }} + {P_{\min }} \le \frac{2}{{{p_j}}}{P_m}.
    \end{cases}
  \end{equation}
  For the feasibility, the constraints on $p_j$, $P_{\max}$ and $P_{\min}$ can be written as
  \begin{subequations}\label{eqn:subs}
    \begin{numcases}{}
      {p_j} \ge 1 - \epsilon,\\
      \frac{\epsilon}{{{p_j}}}\left( {{P_{\max }} - {P_{\min }}} \right) > 0,\label{eqn:sub1}\\
      \left( {1 - \frac{{1 - \epsilon}}{{{p_j}}}} \right){P_{\max }} + \frac{{1 - \epsilon}}{{{p_j}}}{P_{\min }} > 0, \label{eqn:sub2}\\
      {P_{\max }} + {P_{\min }} \le \frac{2}{{{p_j}}}{P_m},
    \end{numcases}
  \end{subequations}
  and the feasible value range of ${P_a}$ can be written as
  \begin{equation}
    \begin{split}
      {P_a} &\le \min \left[ \left( {1 - \frac{{1 - \epsilon}}{{{p_j}}}} \right){P_{\max }} + \frac{{1 - \epsilon}}{{{p_j}}}{P_{\min }},\right.\\
      &\qquad \qquad \qquad \qquad \qquad \quad \left. \frac{\epsilon}{{{p_j}}}\left( {{P_{\max }} - {P_{\min }}} \right) \right]\\
      &=\begin{cases}
        \!\left( {1 \!-\! \frac{{1 - \epsilon}}{{{p_j}}}} \right)\!{P_{\max }} \!+\! \frac{{1 - \epsilon}}{{{p_j}}}{P_{\min }}, &\frac{{{P_{\min }}}}{{{P_{\max }}}} \!\le \!1 \!-\! {p_j},\\
        \frac{\epsilon}{{{p_j}}}\!\left( {{P_{\max }} \!-\! {P_{\min }}} \right),&\frac{{{P_{\min }}}}{{{P_{\max }}}} \!\ge\! 1\! -\! {p_j}.
      \end{cases}
    \end{split}
  \end{equation}
  Besides, according to (\ref{eqn:omega}), to achieve non-zero covert throughput, we should have $R \le {C_f}$. We note that, when ${p_j} > 1 - \epsilon$ and ${p_j} = 1 - \epsilon$, (\ref{eqn:sub1}) and (\ref{eqn:sub2}) always hold for ${P_{\max }} > {P_{\min }} \ge 0$ and ${P_{\max }} > {P_{\min }} > 0$. Hence, (\ref{eqn:subs}) can be rewritten as
  \begin{equation}\label{eqn:constraint1}
    \begin{cases}
      {p_j} > 1 - \epsilon,\\
      {P_{\max }} > {P_{\min }} \ge 0, \\
      {P_{\max }} + {P_{\min }} \le \frac{2}{{{p_j}}}{P_m},
    \end{cases}
  \end{equation}
and
\begin{equation}\label{eqn:constraint2}
  \begin{cases}
    {p_j} = 1 - \epsilon,\\
    {P_{\max }} > {P_{\min }} > 0,\\
    {P_{\max }} + {P_{\min }} \le \frac{2}{{{p_j}}}{P_m}.
  \end{cases}
\end{equation}
\par
Rearranging (\ref{eqn:constraint1}) and (\ref{eqn:constraint2}) completes the proof.
\end{IEEEproof}

  We note that both \textit{Lemma \ref{lemma:feasibility1}} and \textit{Lemma \ref{lemma:feasibility2}} provide the feasible value ranges for the parameters ${P_a}$, $R$, ${p_j}$ ,${P_{\min }}$ and ${P_{\max }}$ in problem (P1). We also note that ${P_{\min }}$ can be zero only if ${p_j} > 1 - \epsilon$.

By using \textit{Lemma \ref{lemma:feasibility2}}, we derive the solution to the optimization problem (P1.2) in the following theorem.
\begin{theorem}\label{sol:Alice}
  From Alice's perspective, the optimal design can be given by
  \begin{equation}
    \begin{cases}
      {P_a}={P_{au}},\\
      R=R_o,
    \end{cases}
  \end{equation}
  where $P_{au}$ is given in \eqref{eq:Pau} and
  \begin{equation}
    R_o=
    \begin{cases}
      C_f,&\Omega_f|_{P_a=P_{au}} \ge \Omega_n|_{P_a=P_{au}},\\
      C_n,&\Omega_f|_{P_a=P_{au}} \le \Omega_n|_{P_a=P_{au}},
    \end{cases}
  \end{equation}
  is the optimal transmission rate.
%  ${C_n} = {\log_2}\left( 1 + {\frac{P_a}{\sigma _b^2 +P_{\max }}} \right)$, ${C_f} = {\log_2}\left( 1 + {\frac{P_a}{\sigma _b^2}} \right)$.
\end{theorem}
\begin{IEEEproof}
  We note that the covertness constraint (S1) and the average power constraint (S2) in the optimization problem (P1) are both independent of $R$. Thus, as per \textit{Proposition \ref{prop:omega}}, we only need to choose the optimal $P_a$ that maximizes $\max(\Omega_f,\Omega_n)$, and then choose the larger one between $\Omega_f$ and $\Omega_n$. This identifies the optimal $R$. For given $p_j$, ${P_{\min}}$, and ${P_{\max}}$, according to the expressions for $\Omega_f$ and $\Omega_n$, they are both strictly monotonically increasing w.r.t. $P_a$. Thus, the optimal $P_a$ is $P_{au}$. When $\Omega_f$ is larger, $C_f$ is optimal, while $C_n$ becomes optimal when $\Omega_n$ is larger.
\end{IEEEproof}

  If Jammer's parameters are given, to maximize the covert throughput, Alice needs to transmit with the maximum feasible power satisfying the covertness constraint (S1). Since Alice's transmission rate is constraint-independent, according to \textit{Proposition \ref{prop:omega}}, the one that maximizes covert throughput (either $C_f$ or $C_n$) should be chosen. Different from the optimal design at Jammer, the optimal design at Alice is unique.

\subsection{Global Optimal Design}
In this subsection, we aim at maximizing the covert throughput from global perspective. In other words, we jointly design the transmission parameters of Alice and Jammer. The optimization problem we face in this case is (P1).
%  Thus, (P1) can be reformulated as
% \begin{align}
%   \text{(P1.3):} \quad & \mathop {\text{maximize  }}\limits_{R,P_a,P_{\min},P_{\max},p_j} \Omega &\notag\\
%   &\text{s.t.} \quad \text{(S1)},\text{(S2)}.&
%   \label{p1.3}
% \end{align}
\par
The first thing we need to discuss is still the feasibility of the optimization problem such that (P1) always has a solution. This can be derived from \textit{Lemmas \ref{lemma:feasibility1}} and \textit{\ref{lemma:feasibility2}}. Also according to \textit{Theorem \ref{theorem:min_xi}}, for any $P_m>0$ and $\epsilon\in(0,\frac{1}{2})$, there exists $P_{\min}$, $P_{\max}$ and $P_a$, satisfying $0 < P_a \le P_{\min}< P_{\max}$, $\frac{1}{2} \left(P_{\min} + P_{\max}\right) \le P_m$, and $\frac{{{P_L}}}{{{P_L} + {P_a}}} \ge 1-\epsilon$. Taking $p_j=\frac{{{P_L}}}{{{P_L} + {P_a}}}$, as in Table \ref{tab:opt_par}, we have $\xi^\ast = p_j \ge 1-\epsilon$, and $\frac{1}{2} p_j \left(P_{\min} + P_{\max}\right) \le P_m$. Hence, there are always feasible parameters that satisfy the constraints (S1) and (S2), which leads to that there is always a feasible solution to (P1).
\par
Since the constraints (S1) and (S2) are independent of $R$, according to \textit{Proposition \ref{prop:omega}}, (P1) can be transformed into
\begin{align}
  \text{(P1.3):} \quad & \mathop {\text{maximize}}\limits_{P_a,P_{\min},P_{\max},p_j} \max(\Omega_f,\Omega_n) &\notag\\
  &\text{s.t.} \quad \text{(S1)},\text{(S2)}.&
  \label{p1.3}
\end{align}
By solving problem (P1.3), the optimal $P_a$, $P_{\min}$, $P_{\max}$, and $p_j$ can be obtained, and then the optimal $R$ can be determined by comparing $\Omega_f$ and $\Omega_n$ based on \textit{Proposition~\ref{prop:omega}}.

We first present the solution to the optimization problem (P1.3) in the following theorem.
\begin{theorem}\label{sol:global1}
  Alice's optimal transmit power $P_a^\ast$, Jammer's optimal transmission probability $p_j^\ast$, the optimal minimum AN power $P_{\min}^\ast$, and the optimal maximum AN power $P_{\max}^\ast$ for maximizing the covert throughput are given by ${P_a^\ast} = \frac{2\epsilon}{1 - {\epsilon^2}}{P_m}$, $p_j^\ast=1-\epsilon$, $P_{\min}^\ast= \frac{2\epsilon}{1 - {\epsilon^2}}{P_m}$, and $P_{\max}^\ast=\frac{2}{1 - {\epsilon^2}}{P_m}$, respectively.
\end{theorem}
\begin{IEEEproof}\label{proof:optimal_pars1}
  According to \textit{Lemma \ref{lemma:feasibility1}}, we have ${P_{\max }} \ge \frac{1}{\epsilon}{P_a}$ for any given $p_j$ and $P_a$. Moreover, $P_a$ achieves the maximum value of $\frac{{2\epsilon}}{{1 - {\epsilon^2}}}{P_m}$ when $p_j=1-\epsilon$, $P_{\min}={P_a}$, and $P_{\max}=\frac{1}{\epsilon}{P_a}$. Thus,
  \begin{equation}
    \begin{split}
      {\Omega _f} &= \left( {1 - {p_j}} \right){\log _2}\left( {1 + \frac{{{P_a}}}{{\sigma _b^2}}} \right)\\
      &\mathop \le \limits^{(a)} \epsilon {\log_2}\left( 1 + \frac{2\epsilon}{1 - {\epsilon^2}}{\frac{P_m}{\sigma _b^2}} \right),
    \end{split}
  \end{equation}
  \begin{equation}
    \begin{split}
      {\Omega _n} &= {\log_2}\left( 1 + {\frac{P_a}{\sigma _b^2 +P_{\max }}} \right) \\
      &\mathop \le \limits^{(b)} {\log_2}\left( 1 + {\frac{P_a}{\sigma _b^2 +\frac{1}{\epsilon}{P_a}}} \right) \\
      &\mathop \le \limits^{(c)} {\log _2}\left( {1 + \frac{2\epsilon {P_m}}{{\left( {1 - {\epsilon^2}} \right){\sigma _b^2} +2{P_m}}}} \right),
    \end{split}
  \end{equation}
  where $(a)$ follows by applying $p_j \ge 1-\epsilon$ and ${P_a} \le \frac{{2\epsilon}}{{1 - {\epsilon^2}}}{P_m}$, $(b)$ follows that $P_{\max} \ge \frac{1}{\epsilon}{P_a}$, and $(c)$ is due to ${P_a} \le \frac{{2\epsilon}}{{1 - {\epsilon^2}}}{P_m}$. Thus, either ${\Omega _f}$ or ${\Omega _n}$ can be the maximum covert throughput when $p_j=1-\epsilon$, $P_{\min}={P_a} = \frac{2\epsilon}{1 - {\epsilon^2}}{P_m}$, and $P_{\max}= \frac{1}{\epsilon}{P_a}=\frac{2}{1 - {\epsilon^2}}{P_m}$.
\end{IEEEproof}

  Intuitively, when $P_{\min}=0$, Jammer has a lower average power compared to the case of $P_{\min}>0$. However, to satisfy the covertness constraint, AN needs to be transmitted with a higher probability. Thus, Jammer's average power consumption becomes higher. Meanwhile, when $P_{\min}=0$, Alice must transmit messages with a lower power, which reduces the covert throughput. Therefore, the optimal value of $P_{\min}$ always satisfies $P_{\min}>0$. When Jammer transmits AN with the minimum feasible probability $1-\epsilon$, Alice's transmit power should be the same as the minimum jamming power $P_{\min}^\ast$. In this way, the maximum covert throughput can be achieved.

We then present the explicit expression for Alice's optimal transmission rate in the following proposition.
\begin{proposition}\label{sol:global2}
  Alice's optimal transmission rate $R^\ast$ for maximizing the covert throughput is given by
  \begin{equation}\label{eqn:opt_r}
    R^\ast=\begin{cases}
      C_f,\quad  \frac{P_m}{\sigma_b^2} \ge {\rho^\ast}\\
      C_n,\quad  \frac{P_m}{\sigma_b^2} \le {\rho^\ast},
    \end{cases}
  \end{equation}
  where $\rho^\ast$ is the only positive solution of
  \begin{equation}
    \epsilon{\ln}\left( 1 + \frac{2\epsilon \rho}{1 - {\epsilon^2}} \right) = {\ln}\left( {1 + \frac{2\epsilon \rho}{{ {1 - {\epsilon^2}}  +2\rho}}} \right).
    \label{eqn:rho}
  \end{equation}
\end{proposition}
\begin{IEEEproof}
  Let $\rho \triangleq \frac{P_m}{\sigma_b^2}$ and $$l(\rho) \triangleq \epsilon{\ln}\left( 1 + \frac{2\epsilon \rho}{1 - {\epsilon^2}} \right)-{\ln}\left( {1 + \frac{2\epsilon \rho}{{ {1 - {\epsilon^2}}  +2\rho}}} \right).$$ According to \textit{Theorem \ref{sol:global1}}, if $l(\rho)>0$, then ${\Omega _f}>{\Omega _n}$, and $R^\ast=C_f$ can be derived. Similarly, $R^\ast=C_n$ if $l(\rho)<0$. Consider the derivative of $l(\rho)$, which is given by $$l^{'}(\rho)=2\epsilon\frac{{4\epsilon{\rho ^2} + 2\epsilon\left( {1 - {\epsilon^2}} \right)\rho  - {{\left( {1 - {\epsilon^2}} \right)}^2}}}{{\left( {1 - {\epsilon^2} + 2\epsilon\rho } \right)\left( {1 - {\epsilon^2} + 2\rho } \right)\left( {1 - \epsilon + 2\rho } \right)}}.$$ Let $l_1(\rho) \triangleq {4\epsilon{\rho ^2} + 2\epsilon\left( {1 - {\epsilon^2}} \right)\rho  - {{\left( {1 - {\epsilon^2}} \right)}^2}}$ and $l_2(\rho) \triangleq {\left( {1 - {\epsilon^2} + 2\epsilon\rho } \right)\left( {1 - {\epsilon^2} + 2\rho } \right)\left( {1 - \epsilon + 2\rho } \right)}$. We note that for any $\epsilon\in(0,\frac{1}{2})$ and $\rho>0$, we have $l_2(\rho)>0$, thus the sign of $l^{'}(\rho)$ is determined by that of $l_1(\rho)$. Let $l_1(\rho)=0$. It has a positive root $\rho_1  = \frac{1}{4}\left( {1 - {\epsilon^2}} \right)\left( {\sqrt {1 + \frac{4}{\epsilon}}  - 1} \right)$ and a negative root $\rho_2  = -\frac{1}{4}\left( {1 - {\epsilon^2}} \right)\left( {\sqrt {1 + \frac{4}{\epsilon}}  + 1} \right)$. Thus, for any $\rho>0$, we have $l_1(\rho)>0$ when $\rho>\rho_1$, and $l_1(\rho)<0$ when $\rho<\rho_1$. This implies that $l^{'}(\rho)<0$ when $\rho<\rho_1$ while $l^{'}(\rho)>0$ when $\rho>\rho_1$. Hence, $l(\rho)$ is decreasing on $\rho\in(0,\rho_1)$ and increasing on $\rho\in(\rho_1,+\infty)$. Since $l(0) = 0$ and there exists $\rho \geq \rho_1$ such that $l(\rho) > 0$. Thus, there exists only one positive solution for $l(\rho)=0$.
\end{IEEEproof}

  Note that $\epsilon \rightarrow 0 \Rightarrow \rho_1 \rightarrow +\infty$. Since $\rho^\ast > \rho_1$, thus $\rho^\ast \rightarrow +\infty$ and $R^\ast=C_n$ always holds. For $\epsilon\in(0,\frac{1}{2})$ and $P_m> \rho^\ast {\sigma_b^2}$, to achieve the maximum covert throughput, Alice should transmit messages at the maximum feasible rate that makes the outage probability $\lambda < 1$ by setting $R=C_f$. Otherwise, Alice should transmit at the maximum feasible rate that makes $\lambda = 0$ by setting $R=C_n$.

\section{Numerical Results}\label{sec:num}
In this section, we present numerical results to verify our analysis and examine the performance of the considered covert communication strategy.

\begin{figure}[t]
  \centering
  %\vspace{-8pt}
  \setlength{\abovecaptionskip}{-3pt}
  \includegraphics[width=1\linewidth]{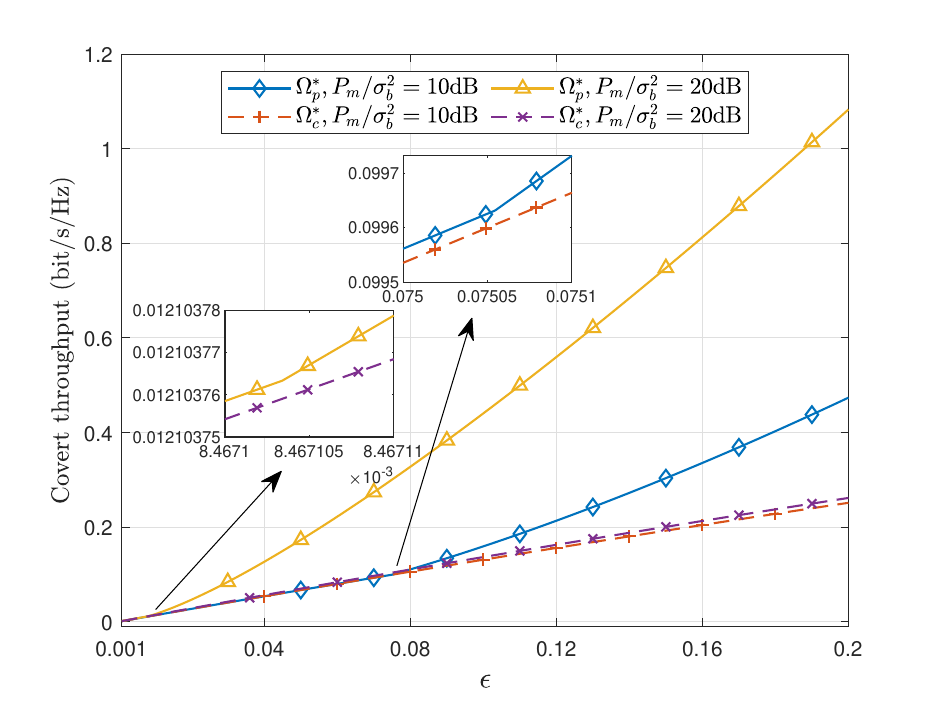}
  \caption{Maximum covert throughput achieved by the probabilistic and continuous jamming strategies $\Omega_p^\ast$ and $\Omega_c^\ast$, versus the covertness constraint $\epsilon$.}\label{fig:throughput_vs_epsilon}
  \vspace{-4pt}
\end{figure}

\begin{figure}[t]
  \centering
  \vspace{-8pt}
  \setlength{\abovecaptionskip}{-3pt}
  \includegraphics[width=1\linewidth]{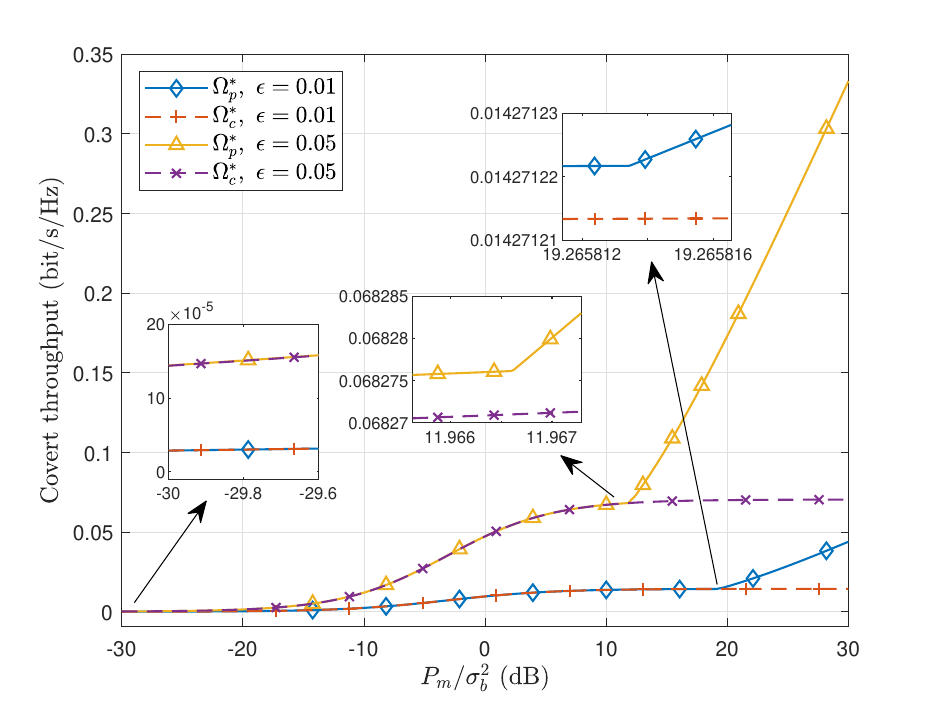}
  \caption{Maximum covert throughput achieved by the probabilistic and continuous jamming strategies $\Omega_p^\ast$ and $\Omega_c^\ast$, versus the ratio of the average jamming power constraint parameter and Bob's receiver noise variance $P_m/\sigma_b^2$.}\label{fig:throughput_vs_Pm}
  \vspace{-4pt}
\end{figure}

We first compare the maximum covert throughput achieved by probabilistic and continuous jamming strategies, represented by $\Omega_p^\ast$ and $\Omega_c^\ast$, respectively. As demonstrated in Figs. \ref{fig:throughput_vs_epsilon}-\ref{fig:throughput_vs_Pm}, we first confirm that with a fixed $\epsilon$ or $P_m/\sigma_b^2$, $\Omega_p^\ast$ is always larger than $\Omega_c^\ast$. This means that the probabilistic jamming can achieve a higher maximum covert throughput than the continuous jamming. We note that the continuous jamming strategy is a special case of the proposed probabilistic jamming strategy with $p_j = 1$. The specific reason for this observation is that the optimal values of $P_a$, $P_{\min}$ and $P_{\max}$ for continuous jamming are $2\epsilon P_m$, $0$ and $2 P_m$, respectively. Thus, according to \textit{Proposition \ref{prop:omega}}, the maximum covert throughput achieved by the continuous jamming is the optimal value of $R$, which is given by $\Omega_c^\ast={\log_2}\left( 1 + {\frac{2\epsilon P_m}{\sigma _b^2 +2 P_m}} \right)$. In contrast, the maximum covert throughput achieved by the probabilistic jamming strategy is $\max\{\Omega_f^\ast,\Omega_n^\ast\}$. We also note that, as $P_m/\sigma_b^2$ increases, the value of $\Omega_c^\ast$ saturates to a constant independent of $P_m/\sigma_b^2$, while the value of $\Omega_f^\ast$ increases continuously. This explains the reason why the probabilistic jamming strategy is capable of significantly outperforming the continuous jamming strategy when $P_m/\sigma_b^2$ is larger than a specific value, as observed from Figs.~\ref{fig:throughput_vs_epsilon}-\ref{fig:throughput_vs_Pm}.

\par
\begin{figure}[t]
  \centering
  %\vspace{-8pt}
  \setlength{\abovecaptionskip}{-3pt}
  \includegraphics[width=1\linewidth]{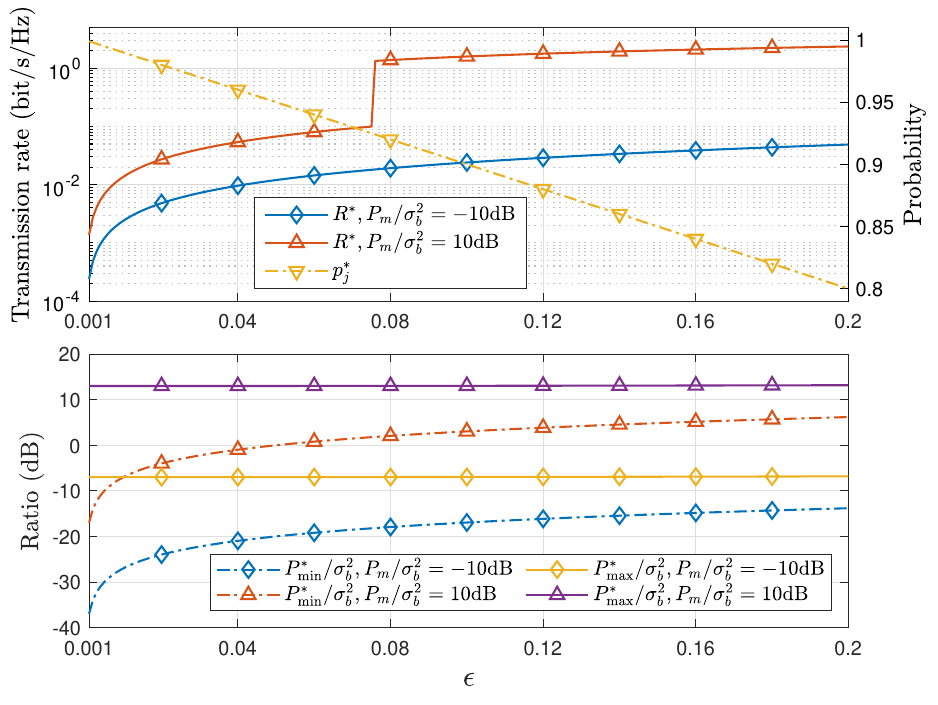}
  \caption{Optimal system parameters $R^\ast$, $p_j^\ast$, ${P_{\min}^\ast}$, and ${P_{\max}^\ast}$ of the probabilistic jamming strategy versus the covertness constraint parameter $\epsilon$.}\label{fig:op_pars_vs_epsilon}
  %\vspace{-4pt}
\end{figure}
\begin{figure}[t]
  \centering
  %\vspace{-8pt}
  \setlength{\abovecaptionskip}{-3pt}
  \includegraphics[width=1\linewidth]{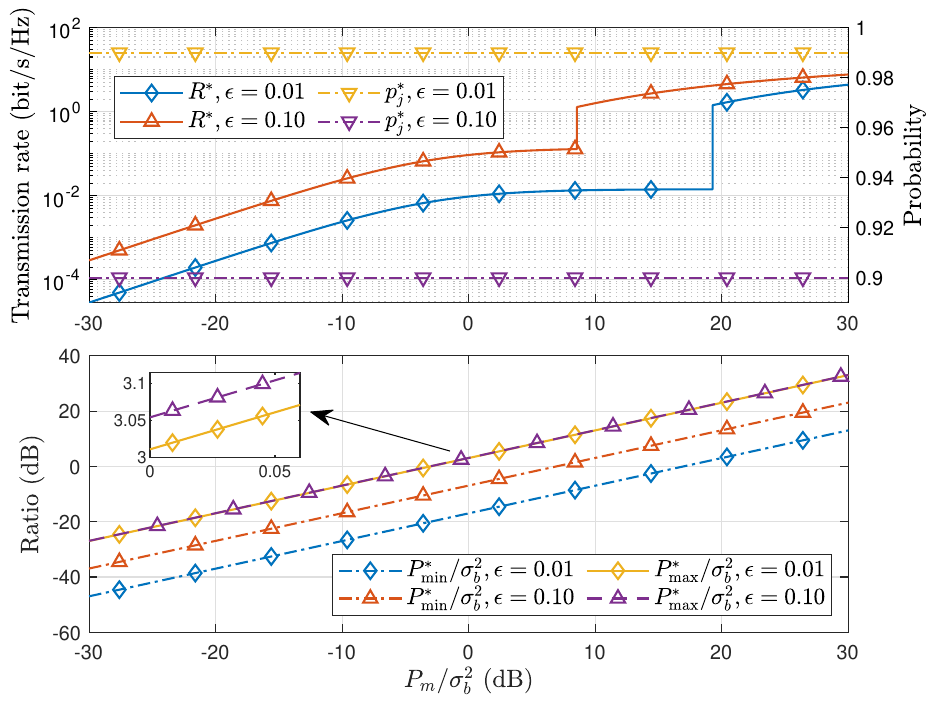}
  \caption{Optimal system parameters $R^\ast$, $p_j^\ast$, ${P_{\min}^\ast}$, and ${P_{\max}^\ast}$ of the probabilistic jamming strategy versus the ratio of the average jamming power constraint parameter and Bob's receiver noise variance $P_m/\sigma_b^2$.}\label{fig:op_pars_vs_Pm}
  %\vspace{-4pt}
\end{figure}

In Figs.~\ref{fig:op_pars_vs_epsilon}-\ref{fig:op_pars_vs_Pm}, we examine the impacts of the covertness constraint parameter $\epsilon$ and Jammer's average power constraint $P_m$ on Alice's optimal transmission rate and transmission power, i.e., $R^\ast$ and $P_a^\ast$, Jammer's optimal transmission probability $p_j^\ast$, the optimal distribution parameters of AN transmit power $(P_{\min}^\ast,P_{\max}^\ast)$. As proved in \textit{Theorem \ref{sol:global1}}, Jammer's optimal minimum jamming power is the same as Alice's optimal transmit power, i.e., $P_a^\ast=P_{\min}^\ast>0$. Hence, we omit $P_a^\ast$ in both plots. First of all, the curves presented in the two figures confirm the correctness of our analysis presented in \textit{Theorem \ref{sol:global1}} and \textit{Proposition \ref{sol:global2}}. Specifically, $p_j^\ast$ decreases with $\epsilon$, which means that Jammer is less likely to transmit AN as the covertness requirement becomes less stringent. In addition, we can observe a sudden jump on Alice's optimal transmission rate $R^\ast$ for the larger $P_m/\sigma_b^2$. This is due to the swap of $R^\ast$ between $C_f$ and $C_n$, as presented in \textit{Proposition \ref{sol:global2}}. Further, we see that $P_{\min}^\ast$ increases with $\epsilon$ while $P_{\max}^\ast$ does not change much with $\epsilon$. The latter observation is mainly due that the considered $\epsilon$ is small. %Intuitively, Jammer sets the minimum jamming power the same as Alice's transmit power to prevent Willie from obtaining the statistical information when Alice is not transmitting, such that to provide a shield for Alice's transmission.

\section{Conclusion}
In this work, we analyzed the detection performance of the adversary and the transmission performance of covert communication with probabilistic jamming on AWGN channels based on a radiometer detector. Adopting the minimum total detection error probability and covert throughput as the metrics, their closed-form expressions were derived, and the scheme to maximize covert throughput was presented. By formulating and solving the optimization problem of maximizing covert throughput subject to a covertness constraint and an average jamming power constraint, the optimal design was derived from the transmitter, the jammer and the global perspectives, respectively. Our numerical results show that the proposed strategy with a probabilistic jammer can achieve a higher covert throughput than that with a continuous jammer under the same covertness and average jamming power constraints. It was revealed that the minimum jamming power should be the same as Alice's transmit power, which depends on the required covertness level and the available average jamming power.

\appendices
\section{Proof of Theorem \ref{theorem:min_xi}}\label{sec:appA}
\begin{figure}[!t]
 \centering
 %\vspace{-5pt}
 %\setlength{\abovecaptionskip}{2pt}
 %\subfigtopskip=2pt
 %\subfigbottomskip=2pt
 %\subfigcapskip=-6pt
 \subfloat[\label{fig:ii1}]{\includegraphics[width=0.5\linewidth]{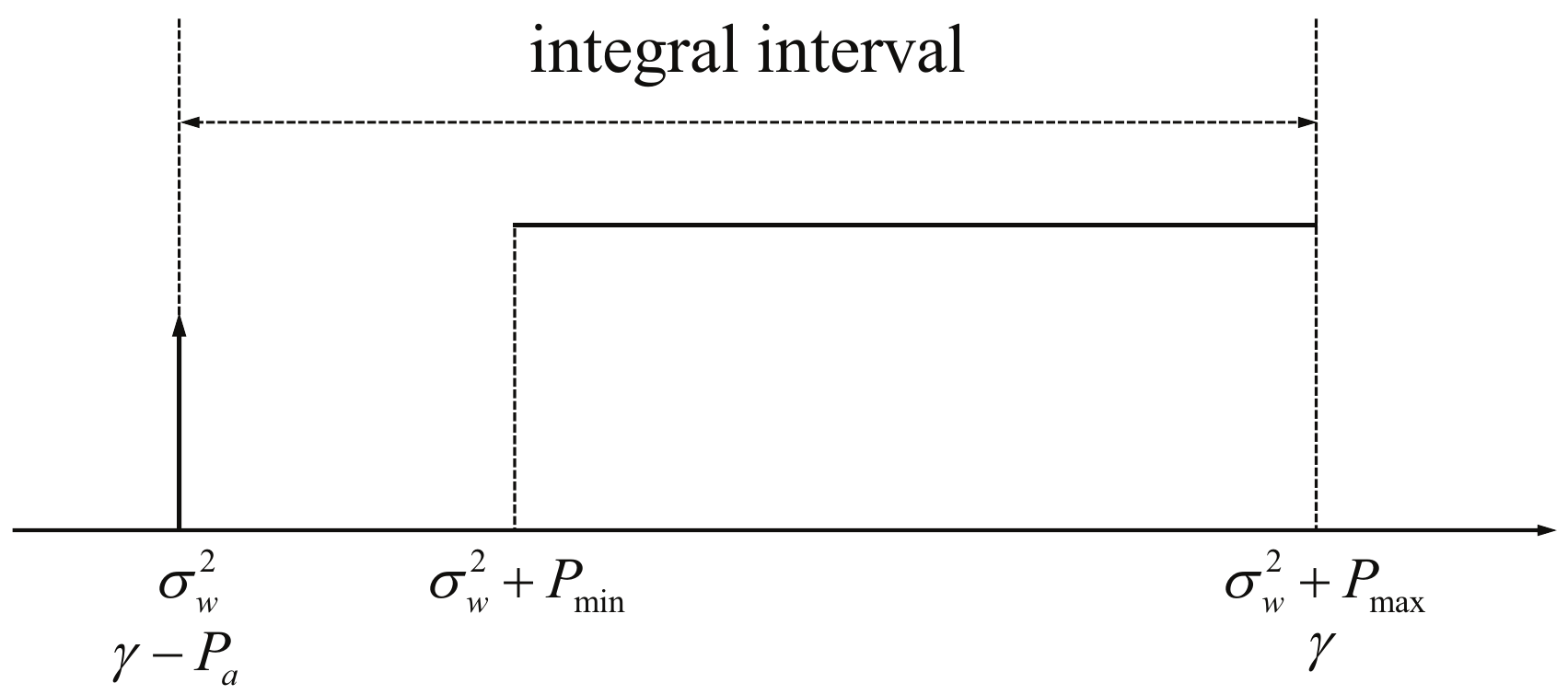}}
 \subfloat[\label{fig:ii10}]{\includegraphics[width=0.5\linewidth]{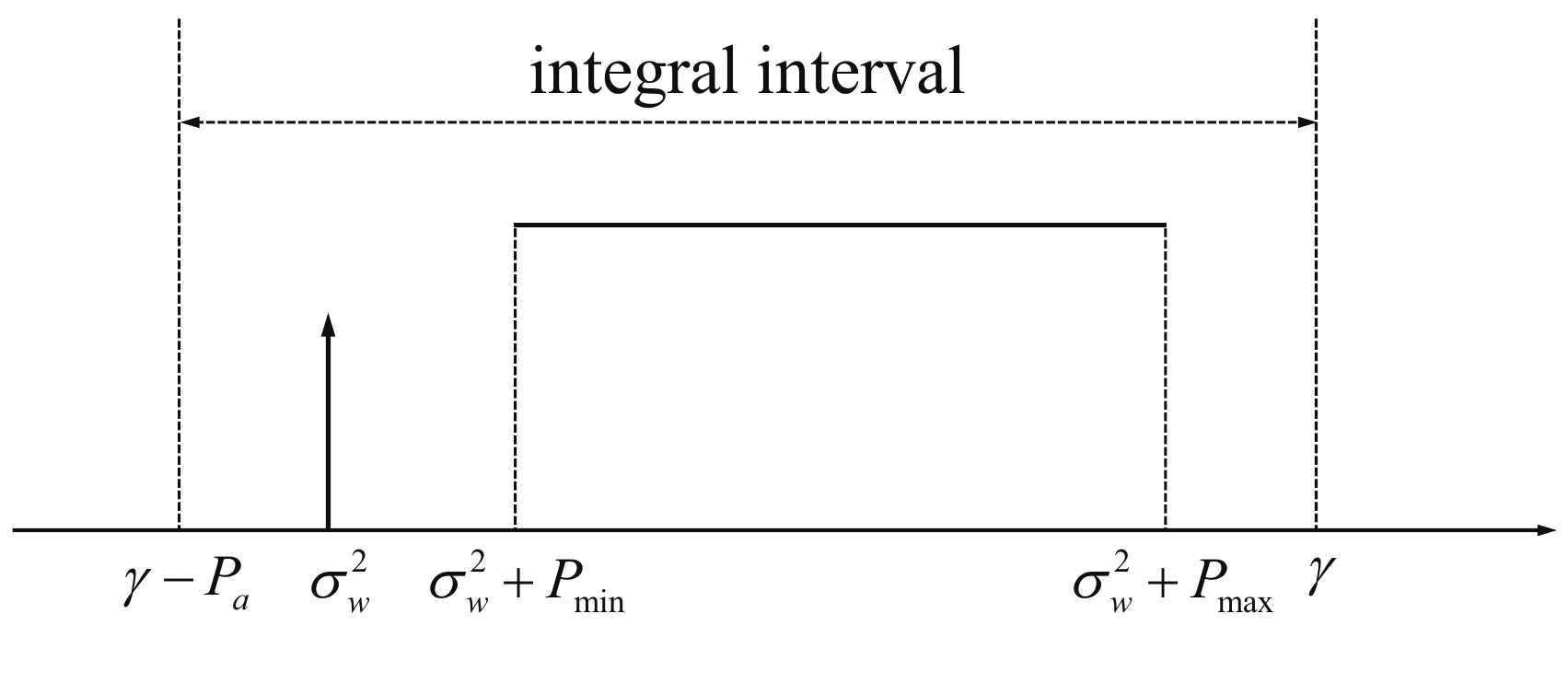}}\\
 \subfloat[\label{fig:ii2}]{\includegraphics[width=0.5\linewidth]{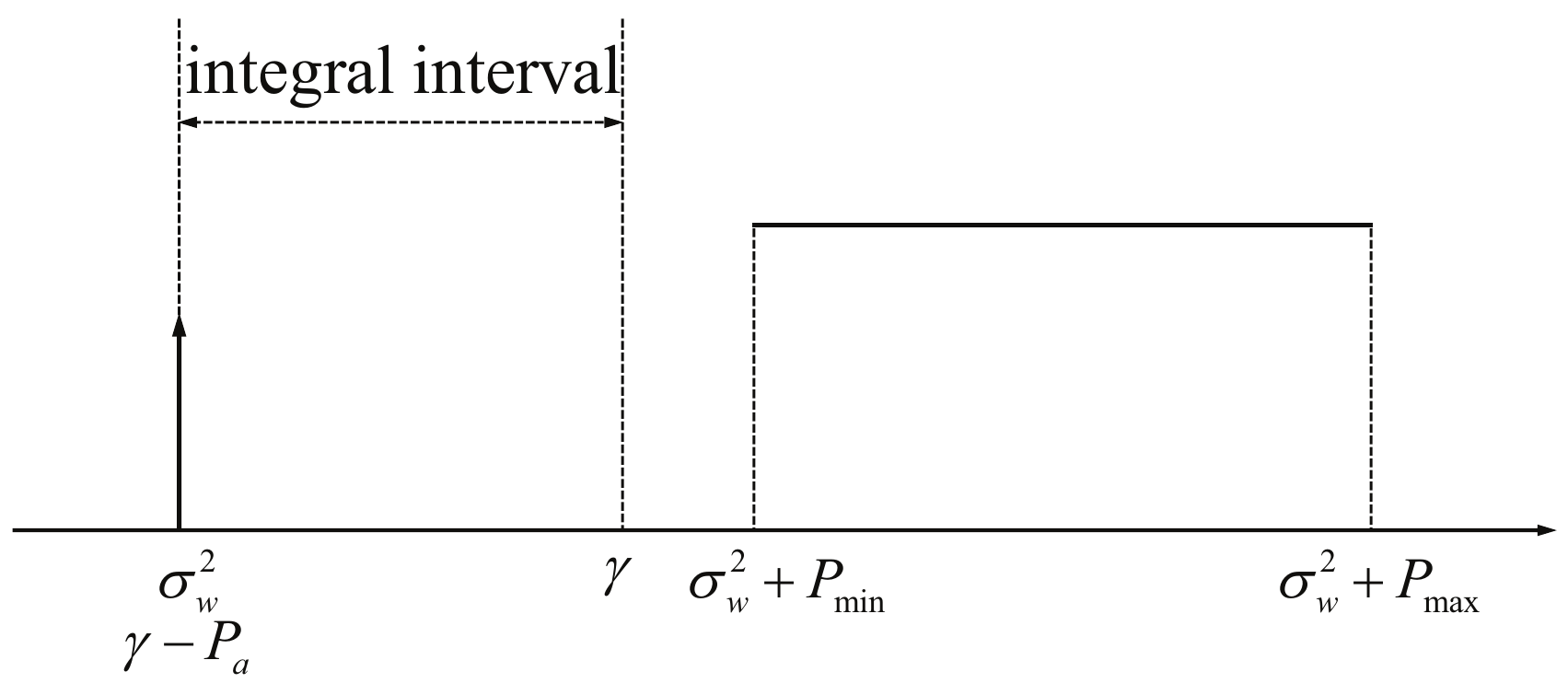}}
 \subfloat[\label{fig:ii3}]{\includegraphics[width=0.5\linewidth]{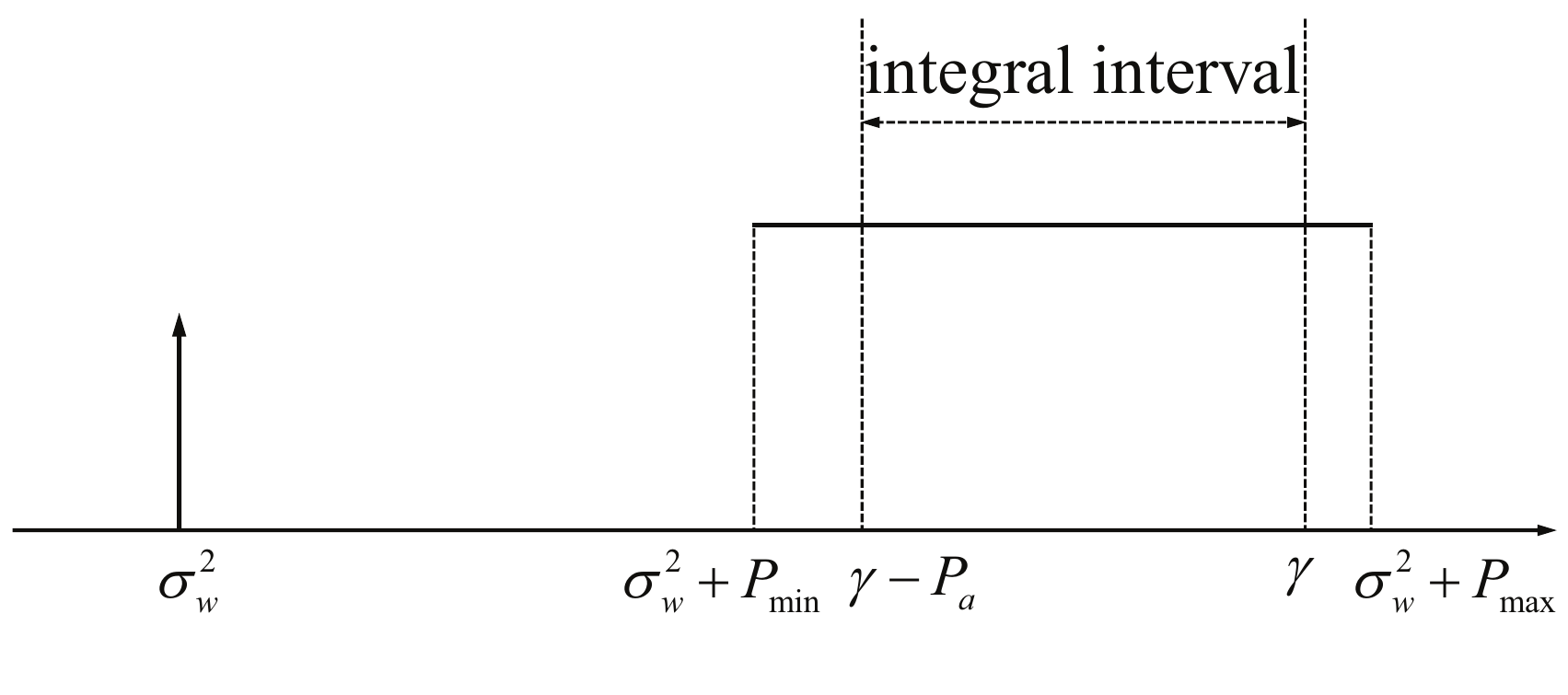}}\\
 \subfloat[\label{fig:ii8}]{\includegraphics[width=0.5\linewidth]{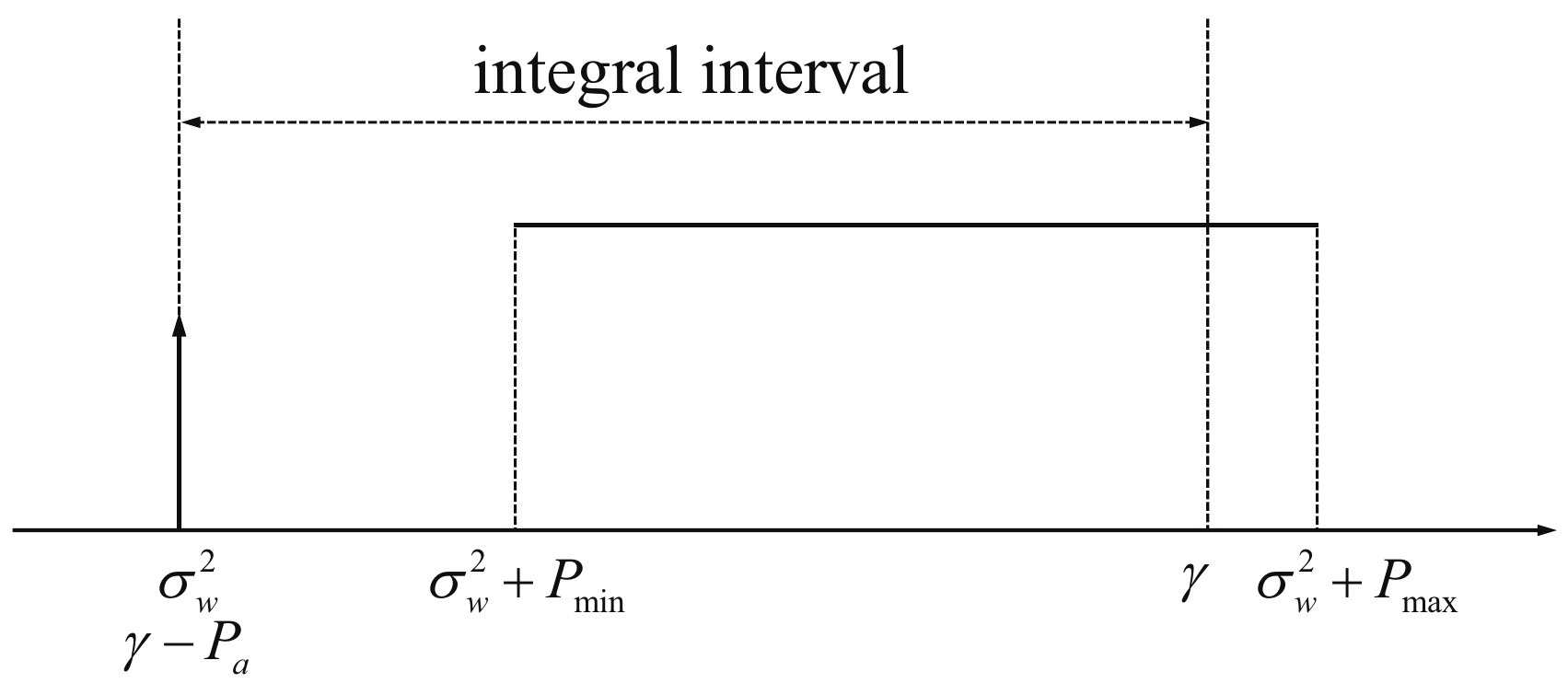}}
 \subfloat[\label{fig:ii9}]{\includegraphics[width=0.5\linewidth]{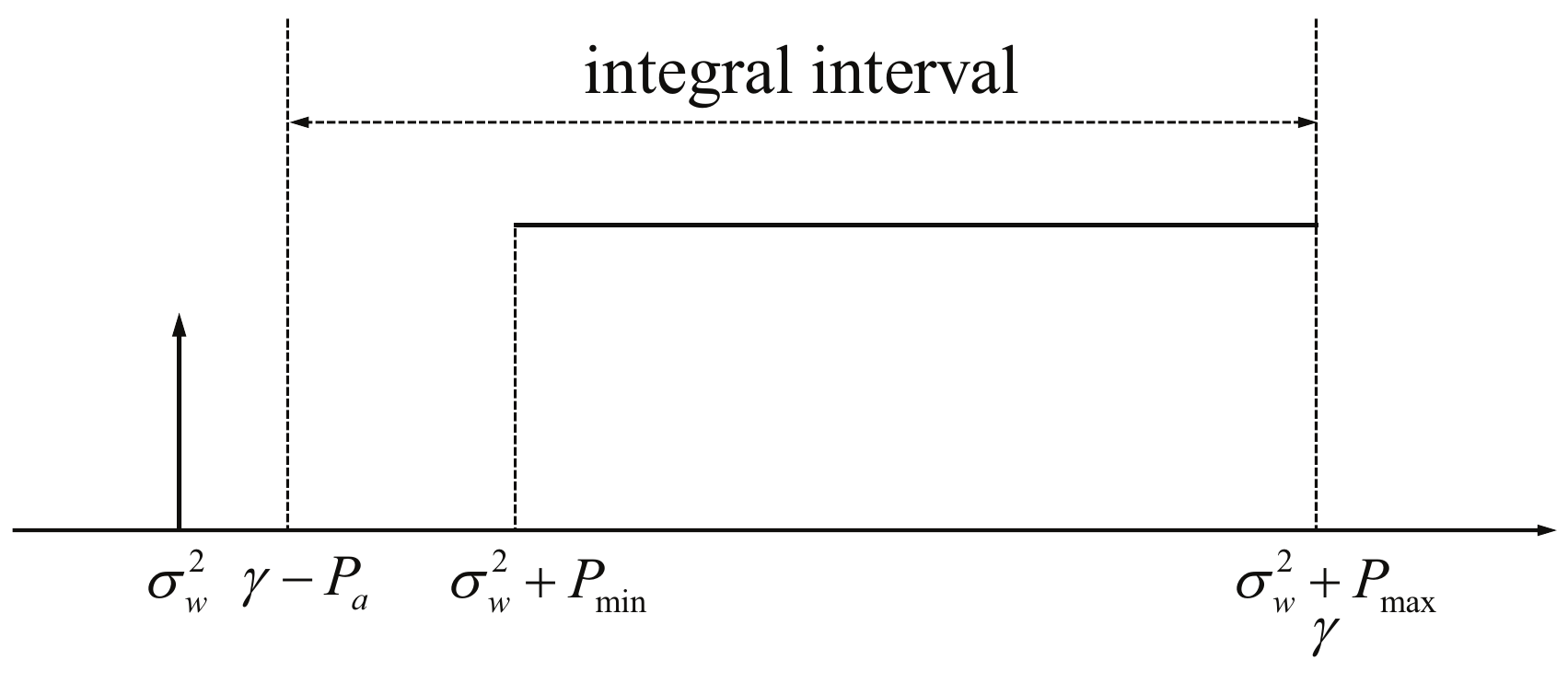}}\\
 \subfloat[\label{fig:ii4}]{\includegraphics[width=0.5\linewidth]{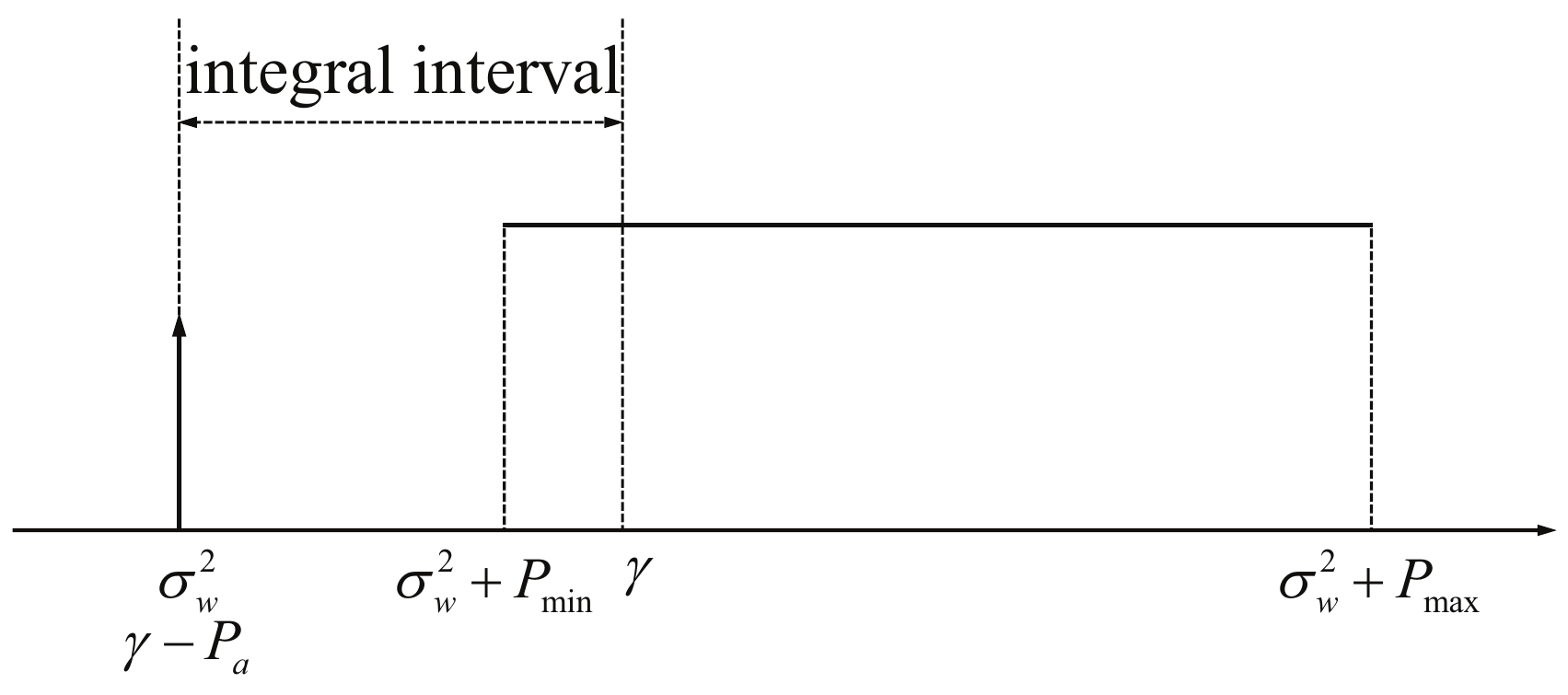}}
 \subfloat[\label{fig:ii5}]{\includegraphics[width=0.5\linewidth]{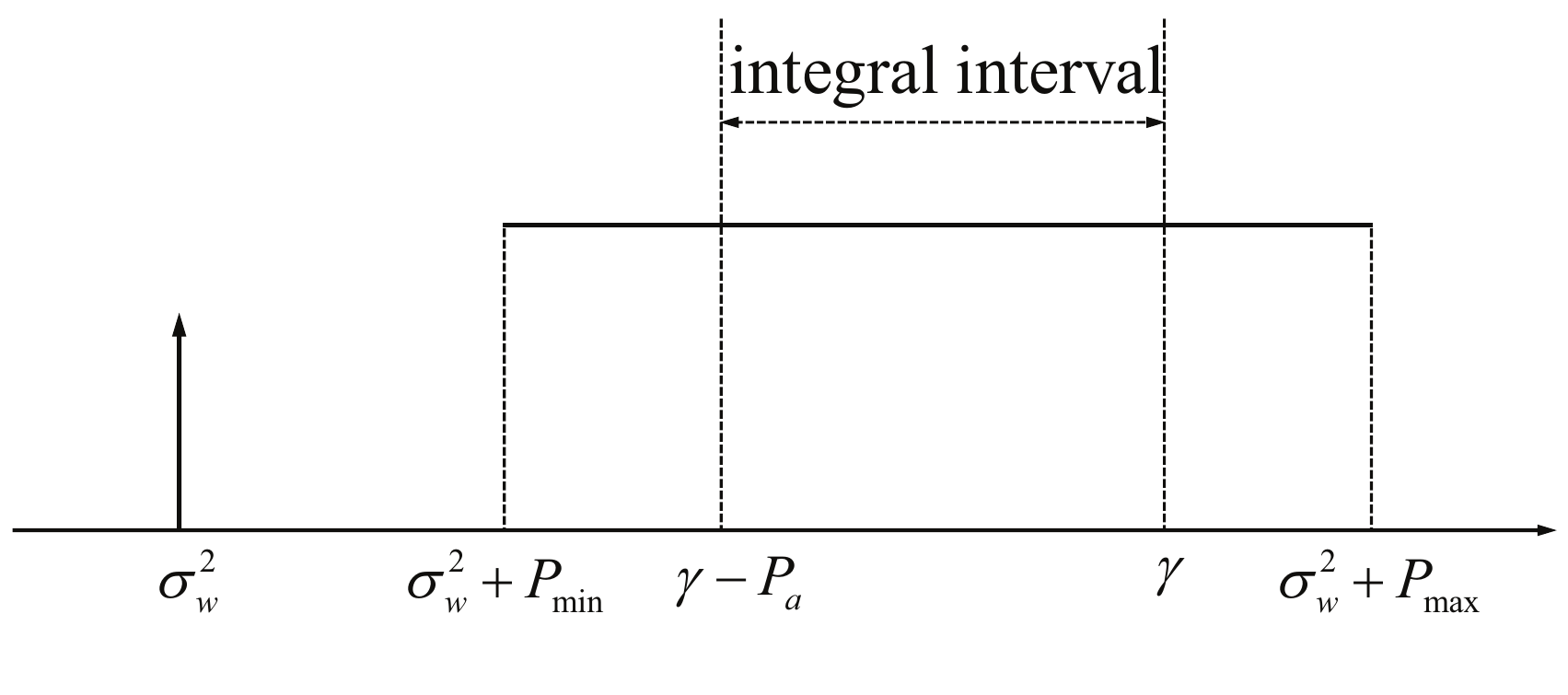}}\\
 \subfloat[\label{fig:ii6}]{\includegraphics[width=0.5\linewidth]{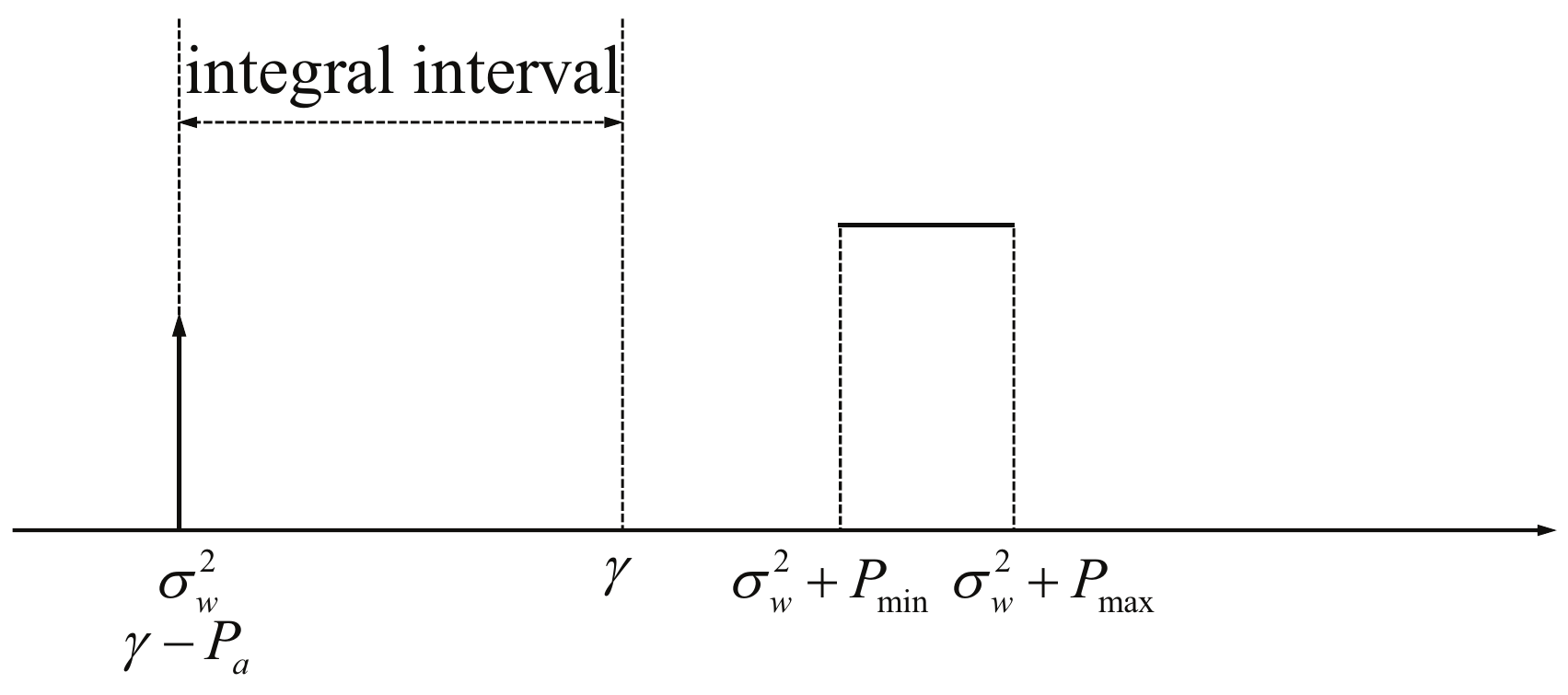}}
 \subfloat[\label{fig:ii7}]{\includegraphics[width=0.5\linewidth]{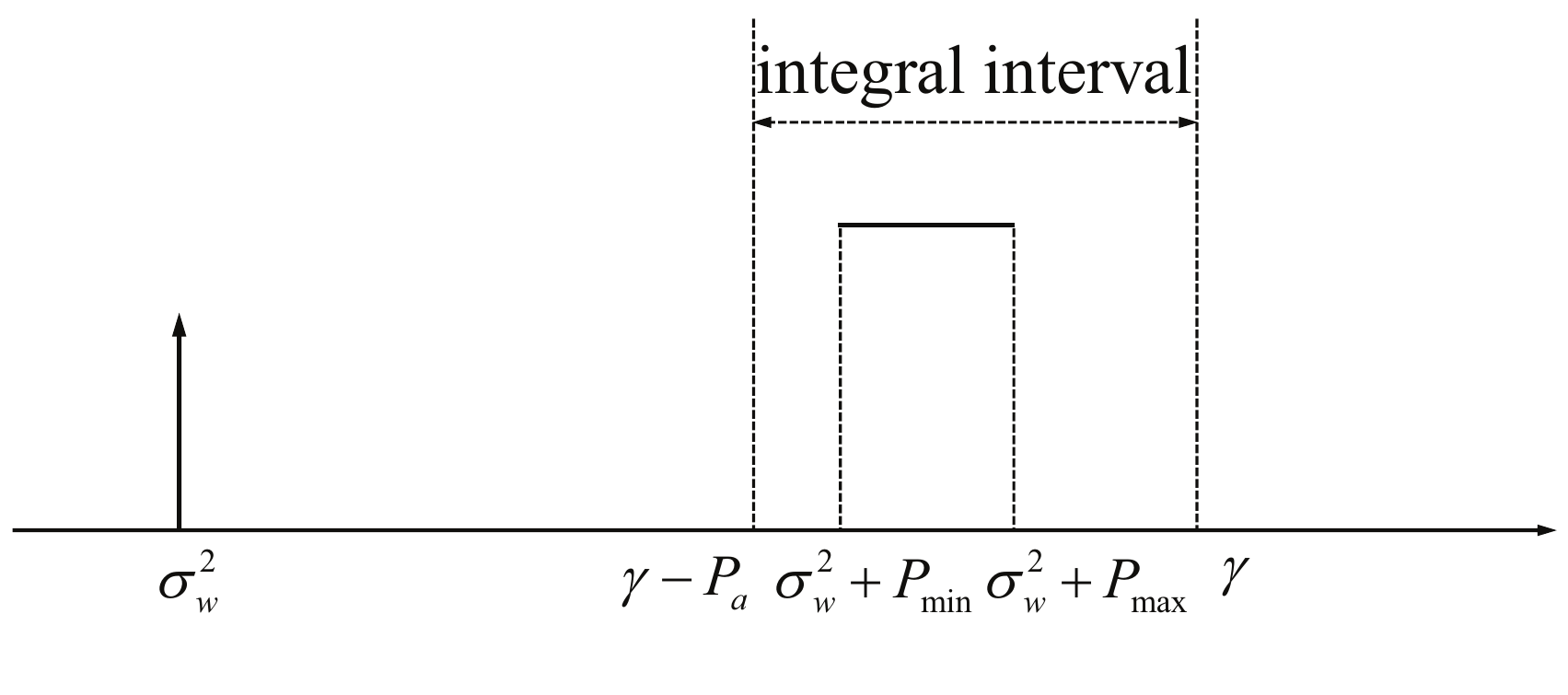}}
 \caption{The integral interval of $\eta$.}
 \label{fig:ii}
 %\vspace{-10pt}
\end{figure}

 Denote the integral interval of $\eta$, the discrete distribution point, the continuous distribution interval and the non-zero distribution interval of ${f_{{P_w}}} ({x|\mathcal{H}_0})$ by $I_i$, $I_d$, $I_c$, and $I_f$, respectively. Specifically, $I_i=[\gamma - P_a,\gamma)$, $I_d=\{\sigma_w^2\}$, $I_c=(\sigma_w^2+{P_{\min}},\sigma_w^2+{P_{\max }})$, and $I_f=[\sigma_w^2,\sigma_w^2+{P_{\max }})$. By using \textit{Lemma \ref{lemma:min_xi}}, we analyze the value of $\gamma$ by case to obtain $\eta^\ast$ and $\gamma^\ast$ in the following.
\begin{enumerate}[]
 \item Case ${P_a} \ge {P_{\max }}$: As shown in Fig.~\ref{fig:ii}(a) and \ref{fig:ii}(b), as long as $I_f \subseteq I_i$, $\eta$ can take the maximum value of $1$. Thus, we have $\gamma^\ast=[\sigma_w^2+ P_{\max }, \sigma_w^2+ P_a]$, $\eta^\ast=1$.
 \item Case ${P_a} \le \min ({P_{\min }},{P_L})$: As shown in Figs.~\ref{fig:ii}(c) and \ref{fig:ii}(d), $I_i$ can cover $I_d$ or a segment of $I_c$. When $I_d \subseteq I_i$, we have $\gamma^\ast=(\sigma_w^2, \sigma_w^2 + {P_a}]$, $\eta^\ast=q_j$. When $I_i$ covers a segment of $I_c$, $\eta$ achieves the maximum value in the case $I_i \subseteq I_c$, we have $\gamma^\ast=[\sigma_w^2+{P_{\min }}+{P_a}, \sigma_w^2 + {P_{\max }}]$, $\eta^\ast=p_j \frac{{P_a}}{{P_L}}$. Thus if $q_j > p_j \frac{{P_a}}{{P_L}} \Rightarrow p_j < \frac{P_L}{P_L + P_a}$, then $I_d \subseteq I_i$. If $p_j > \frac{P_L}{P_L + P_a}$, then $I_i \subseteq I_c$.
 \item Case $\max ({P_{\min }},{P_L}) < {P_a} < {P_{\max }}$: As shown in Fig.~\ref{fig:ii}(e) and \ref{fig:ii}(f), $I_i$ can cover $I_d$ and a segment of $I_c$ or the entire $I_c$. When $I_i$ covers $I_d$ and a segment of $I_c$, then $\gamma^\ast=\sigma_w^2+{P_a}$, $\eta^\ast=1-p_j \frac{{P_{\max }}-{P_a}}{P_L}$. When $I_c \subseteq I_i$, then $\gamma^\ast=[\sigma_w^2 + {P_{\max }},\sigma_w^2+{P_{\min }}+{P_a}]$, $\eta^\ast= p_j$. Hence, if $p_j > 1-p_j \frac{{P_{\max }}-{P_a}}{P_L} \Rightarrow p_j > \frac{{P_L}}{{P_L} +{P_{\max }} - {P_a}}$, then $I_i$ should cover $I_c$. If $p_j < \frac{{P_L}}{{P_L} +{P_{\max }} - {P_a}}$, then $I_i$ should cover $I_d$ and a segment of $I_c$.
 \item Case ${P_{\min }} < {P_a} \le {P_L}$: As shown in Fig.~\ref{fig:ii}(g) and \ref{fig:ii}(h), $I_i$ can cover $I_d$ and a segment of $I_c$ or only a segment of $I_c$. If $I_i$ covers $I_d$ and a segment of $I_c$, we have $\gamma^\ast=\sigma_w^2+{P_a}$, $\eta^\ast=1-p_j \frac{{P_{\max }}-{P_a}}{P_L}$. If $I_i$ covers only a segment of $I_c$, we have $\gamma^\ast=[\sigma_w^2+{P_{\min }}+{P_a}, \sigma_w^2 + {P_{\max }}]$, $\eta^\ast=p_j \frac{{P_a}}{{P_L}}$. As such, if $p_j \frac{{P_a}}{{P_L}} > 1-p_j \frac{{P_{\max }}-{P_a}}{P_L} \Rightarrow p_j > \frac{{P_L}}{{P_{\max }}}$, then $I_i$ should cover only a segment of $I_c$. If $p_j < \frac{{P_L}}{{P_{\max }}}$, then $I_i$ should cover both $I_d$ and a segment of $I_c$.
 \item Case ${P_L} < {P_a} \le {P_{\min }}$: As shown in Fig.~\ref{fig:ii}(i) and \ref{fig:ii}(j), $I_i$ can cover $I_d$ or the entire $I_c$. When $I_d \subseteq I_i$, we have $\gamma^\ast=(\sigma_w^2, \sigma_w^2 + {P_a}]$, $\eta^\ast=q_j$. When $I_c \subseteq I_i$, we have $\gamma^\ast=[\sigma_w^2 + {P_{\max }},\sigma_w^2+{P_{\min }}+{P_a}]$, $\eta^\ast=p_j$. Thus if $q_j > p_j \Rightarrow p_j < \frac{1}{2}$, then $I_d \subseteq I_i$. If $p_j > \frac{1}{2}$, then $I_i \subseteq I_c$.
\end{enumerate}
\par
The above analyses are based on the condition of $0<p_j<1$. It is easy to verify that they are also true for the condition $p_j=1$, except that for the case of ${P_a} \ge {P_{\max }}$, where $\gamma^\ast=[\sigma_w^2+ P_{\max }, \sigma_w^2+ P_{\min }+P_a]$.
\par
Summarizing the above arguments completes the proof.

\section{Proof of Lemma \ref{lemma:feasibility1}}\label{sec:appB}

\begin{figure}[htb]
  \centering
  \includegraphics[width=0.6\linewidth]{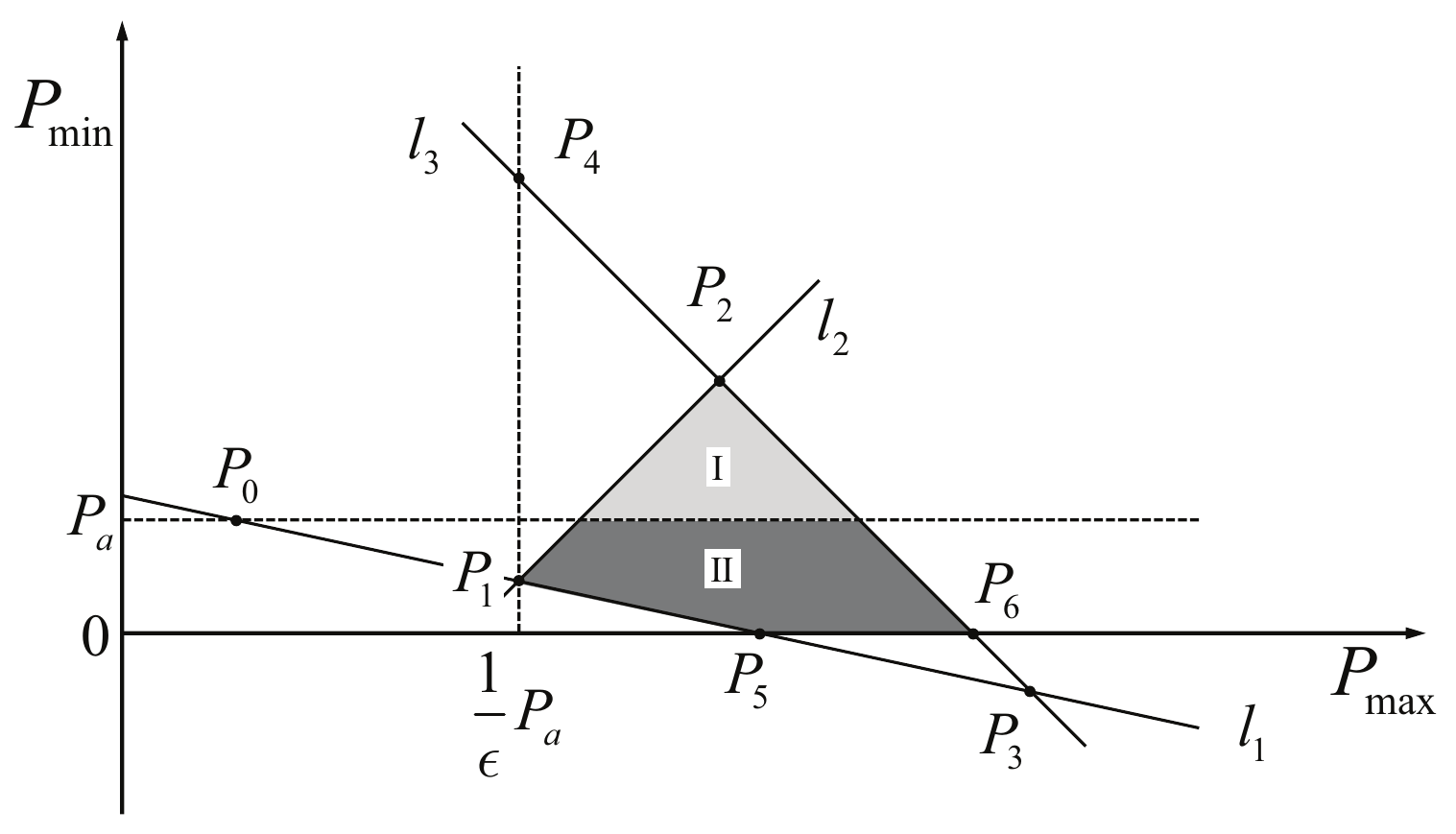}
  \caption{The constrained region for $P_{\min}$ and $P_{\max}$, $\text{I}:{P_{\min }} \ge {P_a}$, $\text{II}:{P_{\min }} \le {P_a}$, ${{l_1}:\frac{{{P_{\min }} - {P_a}}}{{{P_{\max }} - {P_a}}} = 1 - \frac{{{p_j}}}{{1 - \epsilon}}}$, ${l_2}:{P_{\max }} - {P_{\min }} = \frac{{{p_j}}}{\epsilon}{P_a}$, ${l_3}:{P_{\min }} + {P_{\max }} = \frac{2}{{{p_j}}}{P_m}$, ${P_0}:\left({P_a} ,{P_a} \right)$, ${P_1}:\left( {\frac{{{P_a}}}{\epsilon},\frac{{1 - {p_j}}}{\epsilon}{P_a}} \right)$, ${P_2}:\left( {\frac{1}{{{p_j}}}{P_m} + \frac{{{p_j}}}{{2\epsilon}}{P_a},\frac{1}{{{p_j}}}{P_m} - \frac{{{p_j}}}{{2\epsilon}}{P_a}} \right)$, ${P_3}:\left( {\frac{{2\left( {1 - \epsilon} \right){P_m} - p_j^2{P_a}}}{{2\left( {1 - \epsilon} \right){p_j} - p_j^2}},\frac{{2\left( {1 - \epsilon - {p_j}} \right){P_m} + p_j^2{P_a}}}{{2\left( {1 - \epsilon} \right){p_j} - p_j^2}}} \right)$, ${P_4}:\left( {\frac{{{P_a}}}{\epsilon},\frac{2}{{{p_j}}}{P_m} - \frac{{{P_a}}}{\epsilon}} \right)$, ${P_5}:\left( {\frac{{{p_j}}}{{{p_j} + \epsilon - 1}}{P_a},0} \right)$, ${P_6}:\left( {\frac{2}{{{p_j}}}{P_m},0} \right)$.}\label{fig:constraint_region}
\end{figure}

Applying \textit{Corollary \ref{corollary:covertness_constraint}}, for given $p_j$ and $P_a$, the constraints (S1) and (S2) can be written as
\begin{equation}
  \begin{cases}
    \frac{{{P_{\min }} - {P_a}}}{{{P_{\max }} - {P_a}}} \ge 1 - \frac{{{p_j}}}{{1 - \epsilon}}\\
    {P_{\max }} - {P_{\min }} \ge \frac{{{p_j}}}{\epsilon}{P_a} \\
    {P_{\max }} + {P_{\min }} \le \frac{2}{{{p_j}}}{P_m},
  \end{cases}
\end{equation}
which represents the region of a triangle above the horizontal axis, as the region consisting of region I and II shown in Fig. \ref{fig:constraint_region}, where the coordinates of seven interception points $P_0,\ldots,P_6$ are provided in the title.
\par
For feasibility, $P_4$ must be above $P_1$, i.e.,
\begin{equation}
  \frac{2}{{{p_j}}}{P_m} - \frac{1}{\epsilon}{P_a} \ge \frac{{1 - {p_j}}}{\epsilon}{P_a}\label{ineqn:P1P4}
\end{equation}
Solving (\ref{ineqn:P1P4}), we have
\begin{equation}
  {P_a} \le \frac{{2\epsilon}}{{2{p_j} - p_j^2}}{P_m} = \frac{{2\epsilon}}{{1 - {{\left( {1 - {p_j}} \right)}^2}}}{P_m} \le \frac{{2\epsilon}}{{1 - {\epsilon^2}}}{P_m},
\end{equation}
\begin{equation}
  {p_j} \le
  \begin{cases}
    1,& {P_a} \le 2\epsilon{P_m}\\
    1 - \sqrt {1 - 2\epsilon\frac{{{P_m}}}{{{P_a}}}},& 2\epsilon{P_m} \le {P_a} \le \frac{{2\epsilon}}{{1 - {\epsilon^2}}}{P_m}.
  \end{cases}
\end{equation}
\par
Then, referring to Fig. \ref{fig:constraint_region}, the feasible ranges of values for $P_{\min}$ and $P_{\max}$ can be derived.
\par
Besides, according to (\ref{eqn:omega}), to achieve non-zero covert throughput, one should have $R \le {C_f}$.

\section{Proof of Theorem \ref{sol:jammer}}\label{sec:appC}

Denote the minimum value of $\lambda$ by $\lambda^\ast$. For any given feasible ${P_a}$ and $R$, according to the expression of $\lambda$, (\ref{eqn:outage_robability}), we have ${P_r} \ge 0$, and when ${P_{\max }} \le {P_r}$, $\lambda$ achieves the minimum value of 0, i.e., $\lambda^\ast=0$. According to \textit{Lemma \ref{lemma:feasibility1}} and Fig. \ref{fig:constraint_region}, $\min (P_{\max }) = \frac{P_a}{\epsilon}$ always holds. Thus, in the case ${P_r} \ge \frac{P_a}{\epsilon}$, $\lambda^\ast=0$ only if the additional constraint ${P_{\max }} \le {P_r}$ is satisfied.
%Note that when $p_j = 1-\epsilon$, ${P_{\max }} = \frac{P_a}{\epsilon}$ and ${P_{\min }} = {P_a}$, Jammer has the minimum mean power of $\frac{{1 - {\epsilon^2}}}{{2\epsilon}}{P_a}$ and the constraints are always satisfied. Considering that the optimal design is not always unique, e.g., when $R = C_f$, any jamming will cause a outage, no matter what values ${P_{\max }}$ and ${P_{\min }}$ take, there is always $\lambda = p_j$, when the optimal design is not unique, we consider minimizing Jammer's mean power.
  \par
  In the case ${P_r} \le \frac{P_a}{\epsilon}$, ${P_{\max }} \ge {P_r}$ always holds, as per (\ref{eqn:outage_robability}). Depending on the relationship between ${P_{\min }}$ and ${P_r}$, $\lambda$ has two different forms of expressions. Given feasible $p_j$, $\lambda$ is smaller for ${P_{\min }} < {P_r}$ than that for ${P_{\min }} > {P_r}$. Therefore, we need to analyze the feasible range of values of ${P_{\min }}$. Denote the ordinate of $P_3$ by $y_3=\frac{{2\left( {1 - \epsilon - {p_j}} \right){P_m} + p_j^2{P_a}}}{{2\left( {1 - \epsilon} \right){p_j} - p_j^2}}$. Consider the derivative of $y_3$ w.r.t. $p_j$, which is given by
  \begin{equation}
    y_3^{'} = \frac{2 \left( {1 - \epsilon} \right){P_m}}{ {p_j^2{{\left[ {2\left( {1 - \epsilon} \right) - {p_j}} \right]}^2}}}g(p_j),
  \end{equation}
where $g(p_j)={ - \left( {\frac{1}{{1 - \epsilon}} - \frac{{{P_a}}}{{{P_m}}}} \right)p_j^2 + 2{p_j} - 2\left( {1 - \epsilon} \right)}$. Since $\epsilon < \frac{1}{2}$, and ${P_a} \le \frac{{2\epsilon}}{{1 - {\epsilon^2}}}{P_m}$, we have ${\frac{1}{{1 - \epsilon}} - \frac{{{P_a}}}{{{P_m}}}} > 0$. When ${P_a} \le \frac{1}{{2\left( {1 - \epsilon} \right)}}{P_m}$, we have $g(p_j) \le 0$. When ${P_a} > \frac{1}{{2\left( {1 - \epsilon} \right)}}{P_m}$, and $p_{j1} < p_j < p_{j2}$, we have $g(p_j) > 0$, where ${p_{j1,j2}} = \left( {1 - \epsilon} \right)\frac{{{P_m} \pm \sqrt {2\left( {1 - \epsilon} \right){P_a}{P_m} - P_m^2} }}{{{P_m} - \left( {1 - \epsilon} \right){P_a}}}$. Since $2\epsilon < \frac{1}{{2\left( {1 - \epsilon} \right)}}$ always holds for $\epsilon \in \left(0,\frac{1}{2}\right)$, when ${P_a} > \frac{1}{{2\left( {1 - \epsilon} \right)}}{P_m}$, for feasibility, we have $p_j \le 1 - \sqrt {1 - 2\epsilon\frac{{{P_m}}}{{{P_a}}}} < p_{j1}$, and thus $g(p_j) < 0$. Hence, for any given feasible $p_j$, we have $g(p_j) \le 0$, and thus $y_3^{'} \le 0$, i.e., ${P_3}$ moves down as $p_j$ increases. When $p_j=1-\epsilon$, we have $y_3=P_a$, i.e., the minimum feasible values of ${P_{\min }}$ is $P_a$. With some algebraic calculations, when ${P_a} \le 2\epsilon{P_m}$, and $p_j \ge p_{j0}$, we have $y_3 \le 0$, where $p_{j0}=\frac{{{P_m} - \sqrt {P_m^2 - 2\left( {1 - \epsilon} \right){P_m}{P_a}} }}{{{P_a}}}$, i.e., the minimum feasible values of ${P_{\min }}$ is zero. In what follows, we analyze the value of $\lambda$ case by case to obtain the optimal design when ${P_r} \le \frac{1}{\epsilon}{P_a}$.
\begin{enumerate}
  \item Case ${P_a} \le {P_r} \le \frac{1}{\epsilon}{P_a}$: Note that $\min({P_{\min }}) \le {P_r}$ always holds. Thus, for any given feasible $p_j$, there exists feasible $P_{\min}$ that satisfies ${P_{\min }} \le {P_r}$. If ${P_a} \le 2\epsilon{P_m}$, $p_j \ge p_{j0}$, and ${P_{\max }} \ge \frac{{{p_j}}}{{{p_j} + \epsilon - 1}}{P_a}$, we have $\min(P_{\min}) =0$, as per (\ref{eqn:outage_robability}),
\begin{equation}
  \begin{split}
   \lambda  &\mathop \ge \limits^{(a)} {p_j}\frac{{{P_{\max }} - {P_r}}}{{{P_{\max }} - {P_{\min }}}}  \mathop \ge \limits^{(b)}  {p_j}\left( {1\! -\! \frac{{{P_r}}}{{{P_{\max }}}}} \right)\\
   &\mathop \ge \limits^{(c)} {p_j}\left( {1 - \frac{{{P_r}}}{{{P_a}}}} \right) + \left( {1 - \epsilon} \right)\frac{{{P_r}}}{{{P_a}}} \mathop \ge \limits^{(d)} 1 - \epsilon\frac{{{P_r}}}{{{P_a}}},
  \end{split}\label{eqn:optimize_lambda1}
\end{equation}
where $(a)$ follows by using ${P_{\min }} \le {P_r}$, $(b)$ is due to the application of ${P_{\min }} \ge 0$, $(c)$ follows from ${P_{\max }} \ge \frac{{{p_j}}}{{{p_j} + \epsilon - 1}}{P_a}$, and $(d)$ is due to ${p_j} \le 1$.
Otherwise, $\min(P_{\min}) =\frac{{1 - \epsilon - {p_j}}}{{1 - \epsilon}}{P_{\max }} + \frac{{{p_j}}}{{1 - \epsilon}}{P_a}$,
\begin{equation}
  \begin{split}
   \lambda  &\mathop \ge \limits^{(a)} {p_j}\frac{{{P_{\max }} - {P_r}}}{{{P_{\max }} - {P_{\min }}}}  \\
   &\mathop \ge \limits^{(e)}  ( {1 - \epsilon} )\left( {1 - \frac{{{P_r} - {P_a}}}{{{P_{\max }} - {P_a}}}} \right)\\
   &\mathop \ge \limits^{(f)} 1 - \epsilon\frac{{{P_r}}}{{{P_a}}},
  \end{split}\label{eqn:optimize_lambda2}
\end{equation}
where $(e)$ follows that ${P_{\min }} \ge \frac{{1 - \epsilon - {p_j}}}{{1 - \epsilon}}{P_{\max }} + \frac{{{p_j}}}{{1 - \epsilon}}{P_a}$, and $(f)$ is due to ${P_{\max }} \ge \frac{1}{\epsilon}{P_a}$.
Note that if ${P_a} = {P_r}$, we have $\lambda = 0$ when ${P_{\min }}=\frac{{1 - \epsilon - {p_j}}}{{1 - \epsilon}}{P_{\max }} + \frac{{{p_j}}}{{1 - \epsilon}}{P_a}$. i.e., when the point $({P_{\max }},{P_{\min }})$ is on the line $l_1$ as shown in Fig. \ref{fig:constraint_region}. Thus, the optimal design can be given by (\ref{eqn:sol_case3}). If ${P_r} = {P_a}$, then we have $\lambda = 1 - \epsilon\frac{{{P_r}}}{{{P_a}}}$ when ${P_{\max }} = \frac{{P_a}}{\epsilon}$, the optimal design can be given by (\ref{eqn:sol_case2}).
  \item Case ${P_r} < {P_a}$: As per (\ref{eqn:pars_feasible}) and (\ref{eqn:outage_robability}), if ${P_{\min}} \ge {P_r}$, we have $\lambda=p_j \ge 1-\epsilon$, while if ${P_{\min}} \le {P_r}$, we have
  \begin{equation}
    \begin{split}
      \lambda &= {p_j}\frac{{{P_{\max }} - {P_r}}}{{{P_{\max }} - {P_{\min }}}} \\
      &\mathop \ge \limits^{(e)}  ( {1 - \epsilon} ) \left( {1 + \frac{{{P_a} - {P_r}}}{{{P_{\max }} - {P_a}}}} \right) > 1-\epsilon.
    \end{split}
  \end{equation}
 Hence, $\lambda^\ast=1-\epsilon$, and the optimal design can be given by (\ref{eqn:sol_case4}).
\end{enumerate}
\par
Summarizing and converting the conditions w.r.t. $P_r$ to w.r.t. $R$ completes the proof.

\ifCLASSOPTIONcaptionsoff
  \newpage
\fi

\bibliographystyle{IEEEtran}
\bibliography{IEEEfull,reference}
\end{document}